\documentclass[aps,prb,superscriptaddress,showpacs,twocolumn]{revtex4}

\usepackage{epsfig}
\usepackage{subfigure}
\usepackage{times}
\usepackage{amsmath}
\usepackage{amsthm}
\usepackage{bm,amssymb,verbatim}
\usepackage{amsbsy}

\newcommand{\ket}[1]{\vert #1 \rangle}

\newcommand{\ua}{\uparrow}
\newcommand{\da}{\downarrow}
\newcommand{\sgn}{\textrm{sgn}}
\newcommand{\C}{\mathcal{C}}
\newcommand{\e}{\mathcal{E}}

\begin{document}
\title{Majorana fermion exchange in quasi-one-dimensional networks}

\author{David J.~Clarke}
\affiliation{Department of Physics and Astronomy,
University of California, Riverside, CA 92521, USA}

\author{Jay D.~Sau}
\affiliation{Condensed Matter Theory Center and Joint Quantum Institute, Department of Physics,
University of Maryland, College Park, Maryland 20742-4111, USA}
\author{Sumanta Tewari}
\affiliation{Department of Physics and Astronomy, Clemson University, Clemson, SC 29634}

\begin{abstract}
Heterostructures of spin-orbit coupled materials with $s$-wave superconductors are thought to be capable of supporting zero-energy Majorana bound states. Such excitations are known to obey non-Abelian statistics in two dimensions, and are thus relevant to topological quantum computation (TQC). In a one-dimensional system, Majorana states are localized to phase boundaries. In order to bypass the constraints of one-dimension, a wire network may be created, allowing the exchange of Majoranas by way of junctions in the network. Alicea et al. have proposed such a network as a platform for TQC, showing that the Majorana bound states obey non-Abelian exchange statistics even in quasi-one-dimensional systems.\cite{alicea1} Here we show that the particular realization of non-Abelian statistics produced in a Majorana wire network is highly dependent on the local properties of individual wire junctions. For a simply connected network, the possible realizations can be characterized by the chirality of individual junctions. There is in general no requirement for junction chiralities to remain consistent across a wire network. We show how the chiralities of different junctions may be compared experimentally and discuss the implications for TQC in Majorana wire networks.
\end{abstract}
\pacs{ 03.67.Lx, 03.65.Vf, 03.67.Pp, 05.30.Pr}

\maketitle

\section{Introduction}
\label{Sec:Intro}Quantum computation, based on the creative use of the fundamental resources of quantum mechanics, promises exponential speed up of several classically intractable computational problems. However, since quantum states are
extremely susceptible to external perturbations, maintaining their coherence in the presence of environmental interactions is the foremost
challenge in any quantum computation architecture.
In the recently proposed scheme of topological quantum computation (TQC),~\cite{Kitaev,nayak_RevModPhys'08} the environmental decoherence problem
is confronted by encoding quantum information in an intrinsically non-local way, making it essentially immune to any local perturbation due to the environment.
A growing class of theoretically predicted quantum many-body states, characterized by excitations with non-Abelian statistics (non-Abelian anyons),
allow such non-local encoding of quantum information.

Exchange statistics \cite{Wilczek2} is the description of how a many-body wave function transforms under the unitary transposition of any pair of quantum particles. The simplest
 examples of this transformation are those associated with bosonic (multiplication by $1$) and fermionic (multiplication by $-1$) statistics. In $(2+1)$-dimensions, where simple permutation of the coordinates and actual exchange
  of the quantum particles are not necessarily equivalent, the bosonic and fermionic statistics can be generalized to \emph{anyonic} statistics. In Abelian anyonic statistics, the many body wave function can pick up \emph{any} phase between $0$ and $\pi$ under pair-wise exchange of the particles,
   which are now called \emph{anyons}.
 In $(2+1)$-dimensions,
if the many-body ground state wave function happens to be a linear combination of states from a degenerate subspace,
a pair-wise exchange (braiding) of the particles can also unitarily \emph{rotate} the wave function in the ground state subspace. In this case, the braiding statistics can be non-Abelian~\cite{Kitaev,nayak_RevModPhys'08}. The state rotation produced by braiding may be exploited to manipulate the quantum information stored in the ground state manifold, producing quantum gates that may be used for computation.~\cite{Parsa,Zhang,Freedman,Clarke}
 Because of the non-local storage of information within the ground-state subspace, TQC using non-Abelian excitations is intrinsically fault-tolerant. This intrinsic fault tolerance at the hardware level holds considerable promise for the future success of quantum computation.

Non-Abelian quantum systems in the so-called Ising topological class \cite{nayak_RevModPhys'08}
are characterized by topological excitations called Majorana fermions. In some topological superconducting (TS) systems, \cite{schnyder} Majorana fermions arise as non-degenerate zero-energy excitations bound to vortices of the superconducting order parameter.
The second quantized operators, $\gamma_i$, corresponding to the Majorana excitations are Hermitian, $\gamma_i^{\dagger}=\gamma_i$. This is very different from the ordinary fermionic (or bosonic) operators for which $c_i \neq c_i^{\dagger}$. Therefore, each Majorana particle can be regarded as its own anti-particle.\cite{Wilczek-3} Majorana particles have been predicted to occur in some exotic many-body states such as the proposed Pfaffian states in the filling fraction $\nu=5/2$ fractional quantum Hall (FQH) system,~\cite{Moore} spinless chiral $p$-wave superconductors/superfluids, \cite{Read,Ivanov} and non-centrosymmetric superconductors. \cite{Parag, Sato-Fujimoto} Recently, there have been proposals for manufacturing the necessary conditions for Majorana fermions by constructing heterostructure systems in which an $s$-wave superconductor is placed in proximity to the surface of a 3D strong topological insulator (TI) \cite{fu_prl'08} or a semiconductor thin film with Rashba\cite{sau1,Ann} or Dresselhaus\cite{alicea}-type spin-orbit (SO) coupling.

Following this, 1D versions of these systems have been proposed.\cite{roman,Gil,long-PRB} In 1D it is not necessary to have vortex states for Majorana fermions to occur. Instead, Majoranas appear as zero energy modes trapped at the phase boundaries between topologically superconducting (TS) and non-topological superconducting (NTS) phases of the system.
In effect, these would be practical realizations of a 1D lattice model shown earlier by
 Kitaev~\cite{Kitaev1} to contain such Majorana fermion end states. It is somewhat problematic for TQC applications that braiding operations are unavailable in one dimension. Although in principle the Majorana bound states may be moved by gating the system to allow expansion and contraction of the topological regions,\cite{alicea1} there is no room for the Majoranas to pass around one another in 1D. This problem has been resolved by Alicea {\it et al.} \cite{alicea1} by the introduction of a so-called T-junction in a network of quantum wires, which allows one to exchange the end-state Majorana modes using a junction of three quantum wire segments. In this way, their proposal makes a step from a locally one-dimensional system to a globally two-dimensional one, in which braiding is now possible.

 The non-Abelian statistics of
Majorana fermions on a quantum  wire
network on a superconducting substrate
 is not obvious given that the original arguments for
non-Abelian statistics in $p$-wave superconductors relied on the Berry
phase accumulated from taking a Majorana fermion around a vortex.
 \cite{Ivanov}
In their work, Alicea \emph{et. al} have shown\cite{alicea1}
 how non-Abelian
statistics arises in the wire network
by approximately mapping the system to a 1D lattice
model similar to the one considered by Kitaev.~\cite{Kitaev1}
We reproduce this result in a more general setting, and show that the form of the braiding statistics actually implemented during an exchange of Majorana bound states within a wire network is dependent upon local characteristics of the wires and their junctions.

We begin in Sec.~\ref{sec-exchange} by formulating the problem of exchanging
a pair of Majorana fermions in a fashion that is independent of the
underlying wire network. First, we show that given a pair of Majorana operators the signs acquired by them during an adiabatic exchange must be unique and consistent with non-Abelian
statistics. In other words, if $\gamma_1$ and $\gamma_2$
are the 2 Majorana fermions being exchanged, then the result of such an exchange is
 $\gamma_1\rightarrow \lambda\gamma_2$, $\gamma_2\rightarrow -\lambda\gamma_1$ and
$\lambda^2=1$. The result can be described in terms of a
braid matrix written as $U=e^{i\lambda\gamma_1\gamma_2}$.
The braid matrix $U$ associated with a given exchange operation will be shown to be uniquely
determined from the microscopic parameters of the quantum wire network.
 By reversing the trajectory of the exchange operation the sign of $\lambda$
is also reversed.
 In the remainder of the paper, we elucidate the above in the context of a
quantum wire network.
 We show in Sec.~\ref{sec-clockwise-exchange} that the phase acquired by the Majorana fermions when they are exchanged through a junction is determined by a local characteristic of the junction (the junction chirality) that is independent of the exact path taken by the Majoranas, as well as the locations of the TS and NTS regions. We discuss the implications of this characteristic (and particularly the possibility that it will be different for different junctions within the same network) in Sec.~\ref{sec-2-junction} before extending our analysis to multiply connected junctions in Sec.~\ref{sec-multiply}. There we see that in addition to the chirality of junctions there is another representation-invariant quantity relevant to the braiding of Majoranas, i.e. the phase acquired when a Majorana is transported around a loop in the network. This `loop factor' completes the description of the transformations produced by Majorana motion in wire networks. Our analysis appeals only to the general necessity for consistency in the effects of Majorana motion, as well as a few simple assumptions about the network set forth in Sec.~\ref{sec-clockwise-exchange}. With these assumptions satisfied, the results apply to any network of 1D wires supporting Majorana fermions at phase boundaries.

\section{Majorana fermion exchange in the Heisenberg representation}\label{sec-exchange}
The topologically degenerate subspace of states of a system of
topological nanowire segments proximity-coupled to a superconductor may be
manipulated via an adiabatically time-varying
Hamiltonian.  Such operations result in changes of
expectation values of the various observables composed of products
of Majorana fermion operators. These expectation values can be computed
equivalently in both the Schrodinger and the Heisenberg representations.
Therefore the non-Abelian statistics generated by braiding operations can be
studied by analyzing the time-dependent Majorana operator
$\gamma_j(t)=U^\dagger(t)\gamma_j U(t)$ in the Heisenberg
representation. Here $U(t)$ is the unitary time-evolution
operator.

 In this section we will
show that in a general superconducting system an exchange of one pair
 of Majorana fermions $\gamma_1$ and $\gamma_2$ in the
Heisenberg representation is described by a non-Abelian braid matrix
 $U(\tau)=e^{\imath \phi}e^{\pm\pi\gamma_1\gamma_2/4}$, where $\tau$ is a time after the exchange is complete.
 Such a non-Abelian unitary transformation leaves all Majorana fermions
other than $\gamma_{1,2}$ unchanged. The transformation $U$ interchanges
$\gamma_1$ and $\gamma_2$ with a relative $-$ sign such that
$\gamma_1(\tau)=\gamma_2(0)$ and $\gamma_2(\tau)=-\gamma_1(0)$ or
$\gamma_1(\tau)=-\gamma_2(0)$ and $\gamma_2(\tau)=\gamma_1(0)$.
 Below
we show how an exchange of Majorana fermion modes will lead to such a relative
$-$ sign in a general setting.

\subsection{Uniqueness and reversibility of exchange transformation}\label{sec-proof}
We start by describing the process of exchanging
 a pair of Majorana
fermions in terms of the underlying BCS Hamiltonian in
 the Heisenberg representation.
Consider first a Hamiltonian $H_{BCS}$ which has a pair of Majorana
 solutions $\gamma_1$ and $\gamma_2$. Since $\gamma_j$ are zero
 energy Majorana
solutions they commute with the Hamiltonian $([H_{BCS},\gamma_j]=0)$
and they are self-adjoint $(\gamma_j^\dagger=\gamma_j)$.
In order to exchange the pair of Majoranas it is necessary to vary
$H_{BCS}(t)$ adiabatically in time in a certain
time interval $[0,\tau]$. The Hamiltonian is taken to be static before and
after this interval. Thus, one can describe the
states before $t<0$ and after $t>\tau$ by eigenstates of $H_{BCS}(0)$
and $H_{BCS}(\tau)$. Moreover for exchange operations, we will require
that the Hamiltonian at the end of the operation $H_{BCS}(\tau)$
be the same as that at the beginning $H_{BCS}(0)=H_{BCS}(\tau)$.
 Since the evolution was adiabatic,
zero energy Majorana operators at times $t<0$ evolve into zero energy
operators at time $t>\tau$ such that
 $[\gamma_j(t),H_{BCS}(t)]=[\gamma_j(\tau),H_{BCS}(0)]=0$ for $t>\tau$.
If the time variation of the Hamiltonian is such
 that it physically exchanges the positions of the localized Majorana
solutions in the time interval $[0,\tau]$, then it follows that
\begin{equation}
 \gamma_2(\tau)=s_1 \gamma_1(0),\quad\gamma_1(\tau)=s_2 \gamma_2(0)\label{eq:maj0},
\end{equation}
where $s_{1,2}$ are constants.
As a result of the application of the adiabatically time-varying Hamiltonian $H_{BCS}(0<t<\tau)$, the Majorana operators will evolve according to the
 Heisenberg equation of motion
$\dot{\gamma}_j(t)=\imath[H_{BCS}(t),\gamma_j(t)]$. The solution to this equation
can be written formally
 in terms of a unitary operator $U(t)$ as
\begin{equation}
\gamma_j(t)=U^\dagger(t)\gamma_j(0)U(t)\label{eq:maj}
\end{equation}
where $U(t)$ is the time-ordered exponential $U(t)=T e^{-\imath \int_0^t d\tau H_{BCS}(\tau)}$.  From Eq.~\ref{eq:maj}, it is clear that $\gamma_j$ remains Majorana for the entirety
of the time evolution (i.e. $\gamma_j(t)^\dagger=\gamma_j(t)$).
Considering the square of Eq.~\ref{eq:maj} we find that $\gamma_j(\tau)^2=1$ since $\gamma_j(0)^2=1$. Applying
 this relation to the Majorana transformation equation (Eq.~\ref{eq:maj0}), we find that $s_j^2=1$, so $s_j=\pm 1$.
It follows from Eq.~\ref{eq:maj} that $\gamma_{1,2}(\tau)$, and therefore $s_{1,2}$,
can be uniquely determined from the relevant BCS Hamiltonian.
 Using arguments analogous to the ones used above, one can show that
 if the trajectories are reversed i.e.  by replacing
$H_{BCS}(t)$ by $H_{BCS}(\tau-t)$ the values of $s_1$
 and $s_2$ are interchanged.
Thus the exchange operation can be described in terms of a unique
unitary operator $U(\tau)$ which we will refer to as the braid matrix
such that
\begin{align}
&s_1\gamma_1(0)=U(\tau)^\dagger\gamma_2(0)U(\tau)\nonumber\\
&s_2\gamma_2(0)=U(\tau)^\dagger\gamma_1(0)U(\tau)
\end{align}
with the time-reversed braid matrix being described by $\tilde{U}(\tau)=U^\dagger(\tau)$.
\subsection{Non-Abelian statistics}\label{sec-sign}
The above argument only shows that $s_j=\pm 1$. However as discussed above,
non-Abelian statistics implies a relative $-$ sign between the final
Majorana fermions such that $s_1s_2=-1$. In this paragraph, we
show by contradiction that this follows from the conservation of
the fermion parity symmetry obeyed by the BCS Hamiltonian.
Suppose $U=U(\tau)$ is the unitary operator that exchanges a pair of
 Majorana fermions $\gamma_1$ and $\gamma_2$.
 Suppose, for the sake of argument,
  that $\gamma_1,\gamma_2$ do not
pick up a (relative) $-$ sign under $U$.  $U$ then transforms the
 neutral fermion operator $d^\dagger=\gamma_1+\imath\gamma_2$  into $U^\dagger d^\dagger U=\pm\imath d$.  Consider now the action of $U$ on the
ground state of $H_{BCS}=H_{BCS}(0)=H_{BCS}(\tau)$. Since $\gamma_j$
 commute with the
Hamiltonian $H_{BCS}$, so do $d^\dagger$,$d$ and the number
 operator $d^\dagger d$. Thus the ground state can be taken be a
 simultaneous eigenstate of $d^\dagger d$ and $H_{BCS}$. Furthermore if
$|\Psi\rangle$ is a ground state so are $d|\Psi\rangle$ and $d^\dagger|\Psi\rangle$. From here it is straightforward to see that the ground state has
a 2-fold degeneracy: namely the empty state $|0\rangle$ and  $|1\rangle=d^\dagger|0\rangle$. Applying $d^\dagger$ to $U|0\rangle$,
 where $|0\rangle$ is the empty state, we see that
$d^\dagger U|0\rangle=U d\ket{0}=0$, so
\begin{equation}
U|0\rangle=\kappa|1\rangle=\kappa d^\dagger|0\rangle,\label{eq:d}
\end{equation}
 where $\kappa$ is a proportionality constant. The time-dependent BCS Hamiltonian $H_{BCS}(t)$ is symmetric under the unitary
 fermion parity operator $P$ which transforms fermions
 $\psi^\dagger(\bm r)$ as $P:\psi^\dagger(\bm r)\rightarrow -\psi^\dagger(\bm r)$. Thus the initial ground state $|0\rangle$ must be an
 eigenstate of $P$ which has eigenvalues
$\pm 1$ (since $P^2=1$). This is referred to as the ground state having
even or odd parity. Since $P$ commutes with the BCS Hamiltonian $H_{BCS}(t)$ at all times, it must also commute with the unitary time
 evolution $U$. However $d^\dagger$ anti-commutes with the fermion parity operator $P$. This leads to a direct
contradiction with Eq.~\ref{eq:d}, ruling out the possibility $s_1s_2=1$.
Therefore $s_1s_2=-1$, establishing in general the relative $-$ sign
for non-Abelian statistics.

 Thus there are 2 possibilities
for the result of an adiabatic exchange
\begin{align}\label{eq:map}
&\gamma_1\rightarrow \lambda\gamma_2 \textrm{ and } \gamma_2\rightarrow -\lambda\gamma_1
\end{align}
where $\lambda=\pm 1$.
This operation can be compactly represented in the Majorana space as
\begin{equation}
U=e^{\imath \phi}e^{ \pi \lambda\gamma_1\gamma_2/4}
\end{equation}
as claimed. Here the sign in the exponent
 is determined by the path of the adiabatic exchange.
The Abelian phase $\phi$ cannot be determined by considering only the
operator dynamics and requires consideration of the
ground state wave-function. As discussed in the previous subsection,
this braid matrix is uniquely determined by the Hamiltonian.

\section{T-junction exchange}\label{sec-clockwise-exchange}
\begin{figure}
\subfigure{
\includegraphics[trim=4cm 1.5cm 4cm 2.5cm, clip, width=.45\columnwidth]{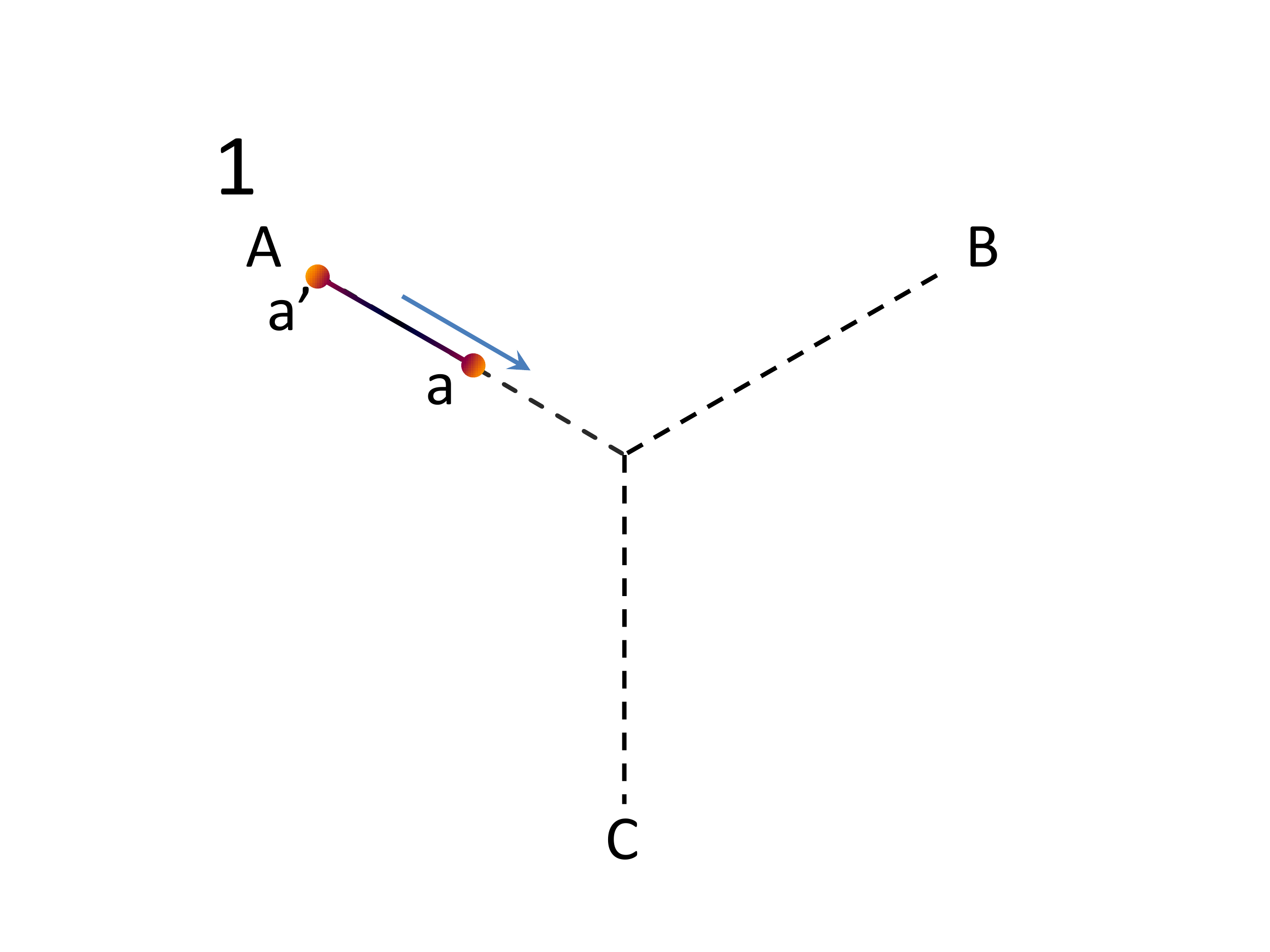}}
\subfigure{
\includegraphics[trim=4cm 1.5cm 4cm 2.5cm, clip, width=.45\columnwidth]{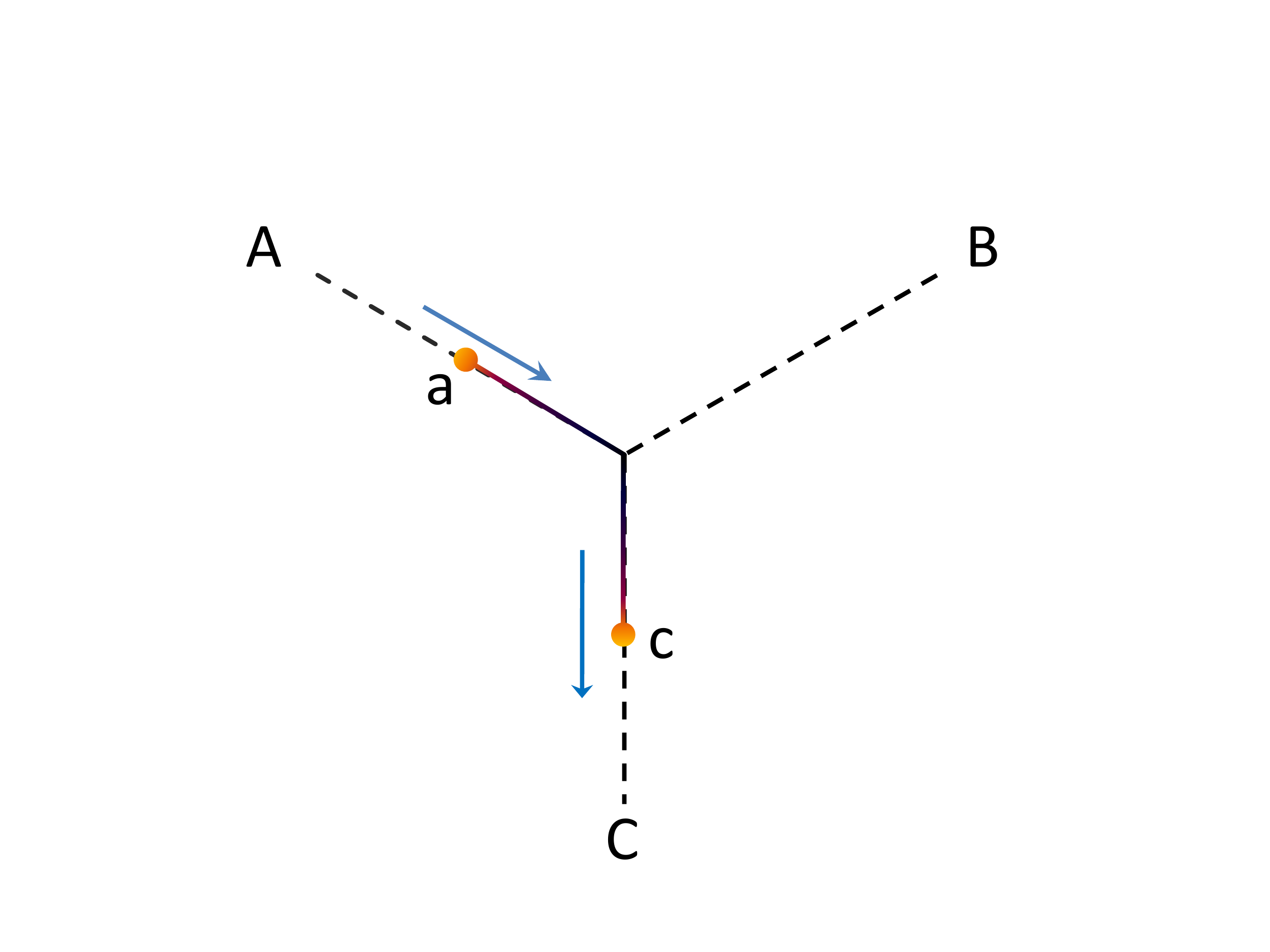}}
\subfigure{
\includegraphics[trim=4cm 1.5cm 4cm 2.5cm, clip, width=.45\columnwidth]{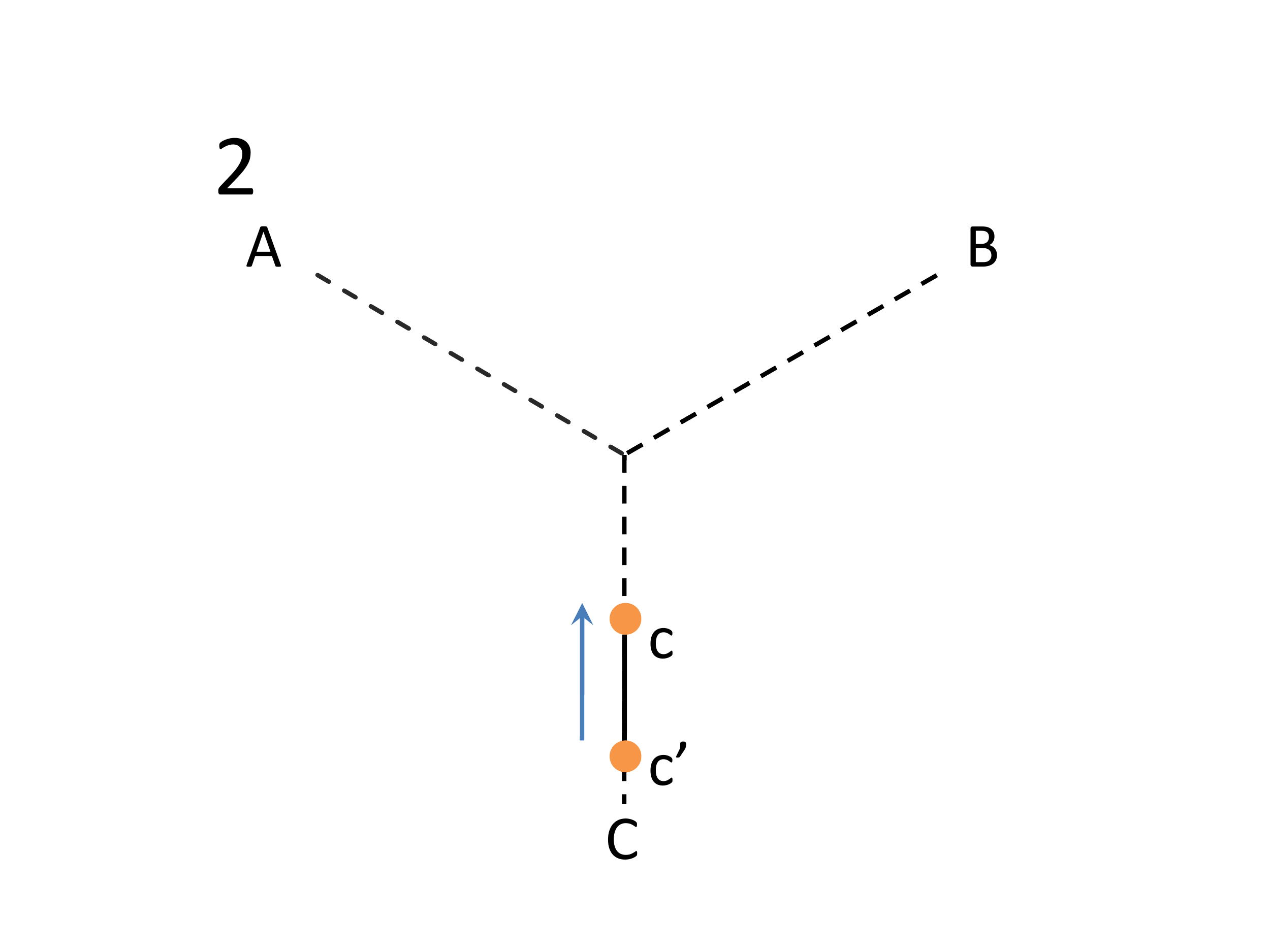}}
\subfigure{
\includegraphics[trim=4cm 1.5cm 4cm 2.5cm, clip, width=.45\columnwidth]{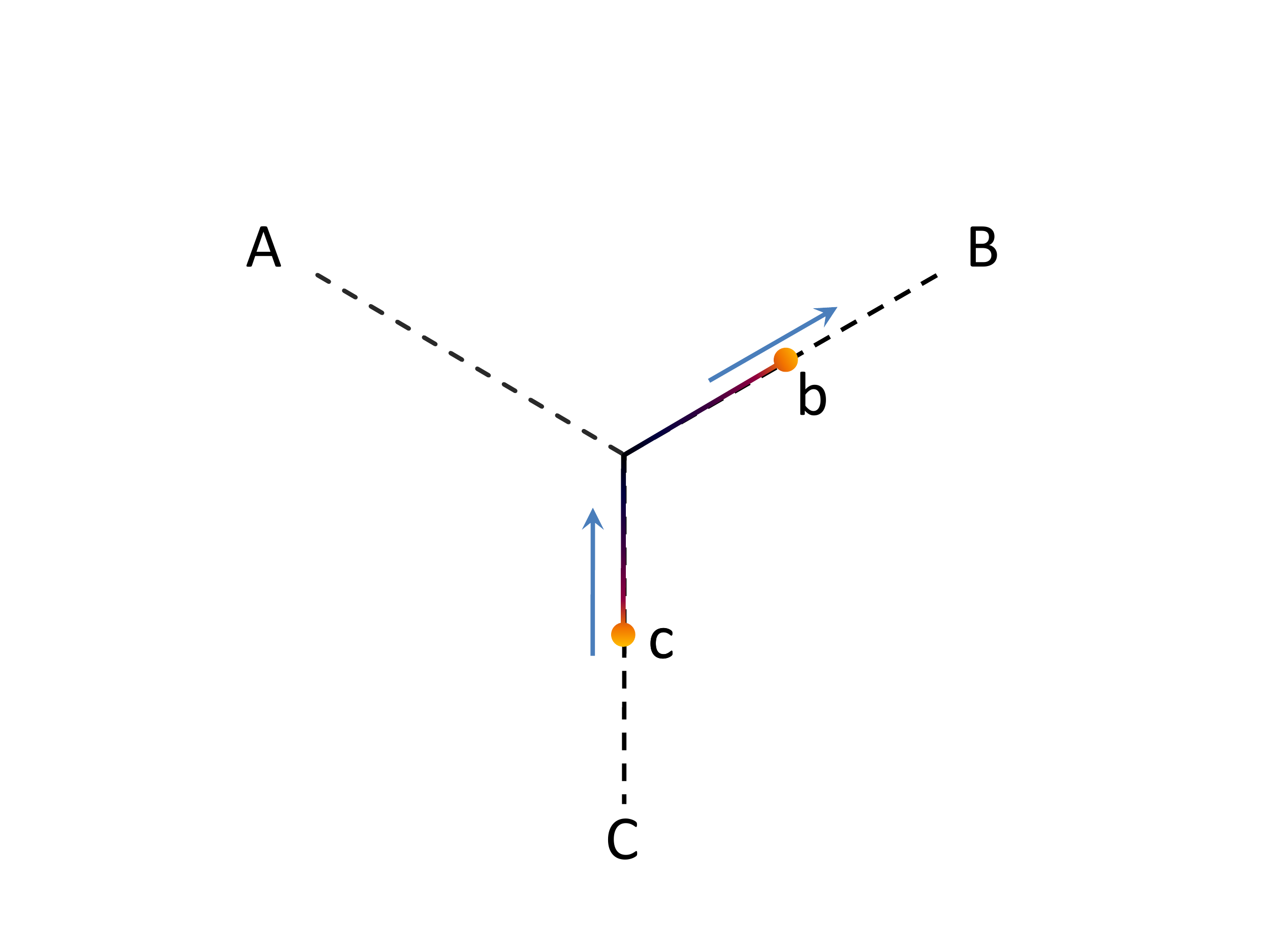}}
\subfigure{
\includegraphics[trim=4cm 1.5cm 4cm 2.5cm, clip, width=.45\columnwidth]{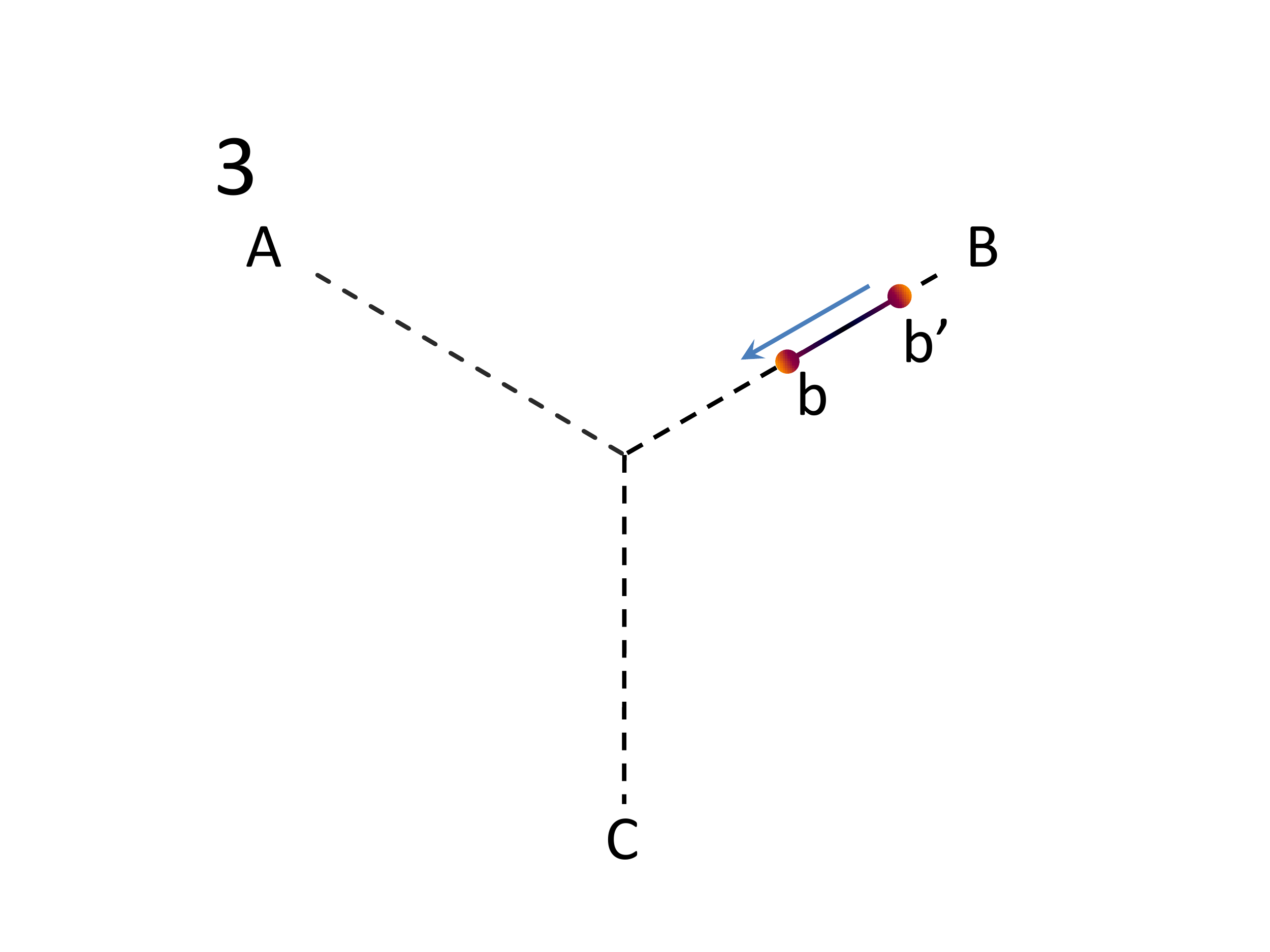}}
\subfigure{
\includegraphics[trim=4cm 1.5cm 4cm 2.5cm, clip, width=.45\columnwidth]{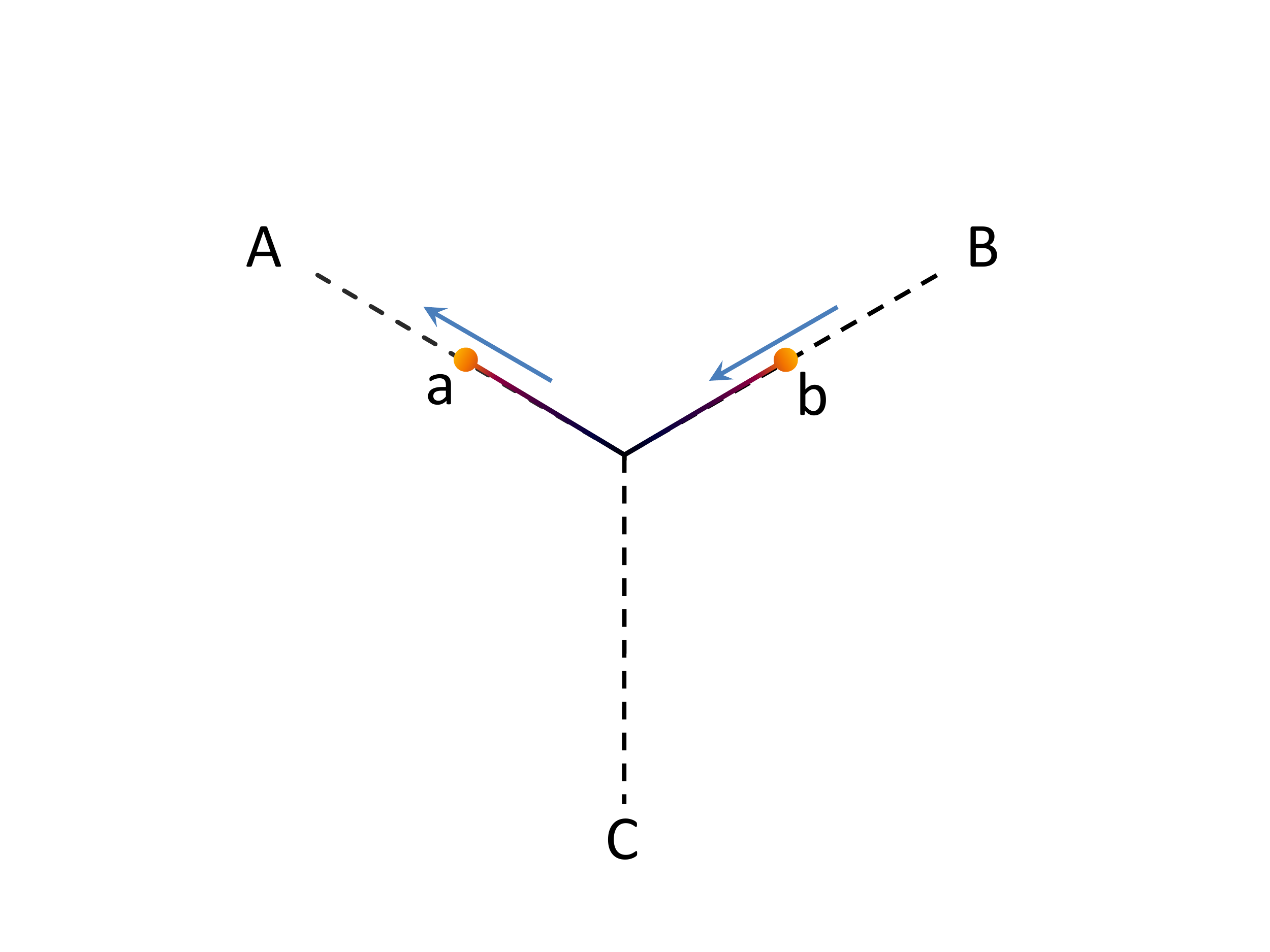}}
\caption{(Color Online) This series of figures shows the interchange of two Majorana bound states at the ends of a (solid) topological region through the activation and deactivation of the wire segments surrounding the junction. Non-topological regions are indicated by dashed lines. At the end of step 3, the system has returned to the configuration shown at the beginning of step 1, but the endpoints have been exchanged.}
\label{T-fig}
\end{figure}
Consider a junction of three wire segments (A, B, and C) as pictured in Fig.~\ref{T-fig}. Initially (at time $t=0$), a portion of the upper left segment (A) is prepared in the topological regime, resulting in two Majorana end states labeled at the points $a'$ and $a$. The procedure to exchange these ends through the junction takes place in three steps, with step $j$ completed at time $t=t_j$, and $t_3=\tau$. For a clockwise exchange gates are activated (1) to extend the topological superconducting phase into segment C; The gates are then deactivated  in segment A, resulting in a contraction of the topological region into segment C. This procedure is repeated (2) to move the topological region into segment B, then (3) back into A. In this process, the two ends of the topological region have been exchanged. As noted above (Eq.~\ref{eq:map}), the resulting transformation must have the form $\{\gamma_{a'}(\tau)=\lambda\gamma_{a}(0),\gamma_{a}(\tau)=-\lambda\gamma_{a'}(0)\}$, where $\lambda=\pm 1$. We shall demonstrate this fact for the most general form of the junction, and show that the effect of an exchange through the junction is the same for Majoranas with the topological and with the non-topological phase in the region between them. In the process, we shall also show that the unitary transformation enacted by an exchange through the junction is independent of the overall sign chosen in the definition of any Majorana state.

In order to determine $\lambda$, we will examine more closely the steps taken in a clockwise exchange. We base this analysis on the following three assumptions about the motion of Majoranas in the wire network:
\begin{tabbing}
\\
1. \=All regions of the wire beside the Majoranas have unique\\
   \>ground states for both the topological and non-topological\\
   \>phase. That is, aside from the Majoranas at the phase\\
   \>boundaries, there are no zero modes in the system.\\\\
2. \>The process of altering the phase boundaries is carried out\\
   \>adiabatically, so that when a topological region is extended\\
   \>or contracted from location $x$ to location $y$, the Majorana\\
   \>zero mode operator at its end is transported according to\\\\
   \>$\gamma_y(t)=\e_{yx}\gamma_x(0)$ for extension\\\\
   \>and\\\\
   \>$\gamma_y(t)=\C_{yx}\gamma_x(0)$ for contraction.\\\\
3.  \>The process is reversible, implying that $\e_{xy}=\C_{yx}$.  \\\\
\end{tabbing}
Without these assumptions, it is impossible to consistently perform braiding operations through the junction. For consistency in expanding within the same segment, we require that
\begin{equation}\label{eq-expansion-consistency}
\e_{zx}=\e_{zy}\e_{yx}
\end{equation}
so long as either points $x$ and $y$, or points $y$ and $z$ are in the same segment.
Likewise, the requirement that Majorana fermions remain properly normalized leads to $\e_{xy}=\pm1$ and $\C_{xy}=\pm1$.

Note that Assumption 1 does not preclude the possibility of localized fermion bound states, but requires that such states are affected deterministically by the passage of a Majorana. What is more, it requires that the effect of passing a Majorana over such localized states is the same independent of whether a topological region is being contracted or expanded. With Assumption 2, this is equivalent to the condition that two Majoranas brought together by the contraction of a single topological (or non-topological) region will be the either in the ground state or the excited state independent of the location at which they are brought together.

If the parity of a short topological region at $x$, as determined by the fusion channel $\imath\gamma_{x-\epsilon}\gamma_{x}$ of the two Majoranas at its end points, does not match the parity of the ground state when that region is eliminated, a quasiparticle will be left behind by the elimination of the topological region.
This allows us to define the local vacuum channel $v_x$ for two Majoranas by the form of the Hamiltonian
\begin{equation}\label{ham}
H=-i\Gamma^{(n,t)}(x,\epsilon) \gamma_{x-\epsilon}\gamma_{x}
\end{equation}
when they are brought very close together, where $\Gamma^{(n,t)}$ is the coupling constant for two Majoranas with a topological (t) or non-topological (n) region between them. We set
\begin{equation}
v^{(n,t)}_x=\lim_{\epsilon\rightarrow0}\sgn\left(\Gamma^{(n,t)}(x,\epsilon)\right),
\end{equation}
where $x-\epsilon$ is always further from the junction than $x$.
Consider the situation in which two Majoranas at points $x_-$ and $x$ are endpoints of the same small topological region and are transported to points $y_-$ and $y$ closer to the junction. ($x_-$ indicates a point infinitesimally further from the junction than $x$
) First the right side of the region is expanded from $x$ to $y$, then the left side is contracted from $x_-$ to $y_-$. Assumptions 1 and 3 together imply that if the system started in a state such that elimination of the topological region would leave the system in an excited state when the elimination happens at $x$, the same must be true if the region is eliminated at $y$. Likewise, if the system began in the ground state of the Majorana interaction Hamiltonian (\ref{ham}) at $x$, it must end in the ground state of the corresponding Hamiltonian at $y$. Therefore,
\begin{eqnarray}
\C_{y_-x_-}=v^t_xv^t_y\C_{xy}
\end{eqnarray}
if points $x$ and $y$ are on the same segment, and
\begin{equation}\label{eq-vac-consistency}
\C_{yx_-}=-v^t_xv^t_y\C_{xy_-}
\end{equation}
if points $x$ and $y$ are on different segments. The minus sign here occurs due to the fact that $\gamma_y$ and $\gamma_{y_-}$ anti-commute, and $y_-$ is further from the junction than $y$. If $x$ and $y$ are endpoints of \emph{different} topological regions, then the same arguments lead to
\begin{equation}
\C_{x_-y_-}=v^n_xv^n_y\C_{yx}
\end{equation}
if $x$ and $y$ are on the same segment and
\begin{equation}\label{eq-vac-consistency2}
\C_{y_-x}=-v^n_xv^n_y\C_{x_-y}
\end{equation}
if $x$ and $y$ are on different segments.
\subsection{Exchange of Majorana endpoints of a single Topological region}
At the first step of the exchange process, the Majoranas are moved from segment $A$ into segment $C$ by first expanding the topological region to move the Majorana that begins nearest the junction (at position $a$) through it to position $c'$ in segment $C$ and then contracting the back end of the topological region through the junction to bring the second Majorana from $a'$ to $c$ (Fig.~\ref{T-fig}). In terms of the Majorana operators, we have
\begin{eqnarray}
\gamma_{c'}(t_1)=\e_{c'a}\gamma_{a}(0)\nonumber\\
\gamma_{c}(t_1)=\C_{ca'}\gamma_{a'}(0).
\end{eqnarray}
At the next step of the process, we use the same procedure to move from segment $C$ to segment $B$, leading to
\begin{eqnarray}
\gamma_{b'}(t_2)=\e_{b'c}\gamma_{c}(t_1)\nonumber\\
\gamma_{b}(t_2)=\C_{bc'}\gamma_{c'}(t_1).
\end{eqnarray}
Finally, the Majoranas are moved back into segment $A$, with
\begin{eqnarray}
\gamma_{a'}(t_3)=\e_{a'b}\gamma_{b}(t_2)\nonumber\\
\gamma_{a}(t_3)=\C_{ab'}\gamma_{b'}(t_2).
\end{eqnarray}
In total, we have that
\begin{eqnarray}
\gamma_{a'}(\tau)=\e_{a'b}\C_{bc'}\e_{c'a}\gamma_{a}(0)\nonumber\\
\gamma_{a}(\tau)=\C_{ab'}\e_{b'c}\C_{ca'}\gamma_{a'}(0).
\end{eqnarray}
Using Eq.~(\ref{eq-expansion-consistency}), we can reduce this to
\begin{eqnarray}\label{eq-exchange-t}
\gamma_{a'}(\tau)&=&\e_{a'a_-}\e_{a_-b}\C_{bc}\e_{ca}\gamma_{a}(0)\nonumber\\
 &=&\C_{a_-a'}v^t_a\chi^t\gamma_{a}(0)\nonumber\\
\gamma_{a}(\tau)&=&\C_{ab}\e_{bc}\C_{ca_-}\C_{a_-a'}\gamma_{a'}(0)\nonumber\\
&=&-\C_{a_-a'}v^t_a\chi^t\gamma_{a'}(0),
\end{eqnarray}
where we have defined
\begin{equation}
\chi^t=v^t_a \e_{a_-b}\C_{bc}\e_{ca}
\end{equation}
and used Eq.~(\ref{eq-vac-consistency}) to permute the indices in the second equation.
Note that $\chi^t$ is defined uniquely for the junction, since
\begin{eqnarray}\label{eq-chi-consistent}
v^t_a\e_{a_-b}\C_{bc}\e_{ca}=v^t_b\e_{b_-c}\C_{ca}\e_{ab}=v^t_c\e_{c_-a}\C_{ab}\e_{bc},\nonumber
\end{eqnarray}
which can be shown using Assumption 2 and Eq.~(\ref{eq-vac-consistency}). Furthermore, due to Eq.~(\ref{eq-expansion-consistency}) $\chi^t$ is independent of the locations of points $a$, $b$ and $c$. It is only important to the definition of $\chi^t$ that each of these three points be on a different one of the segments connected by the junction.

\begin{figure}
\subfigure{
\includegraphics[trim=4cm 1.5cm 4cm 2.5cm, clip, width=.45\columnwidth]{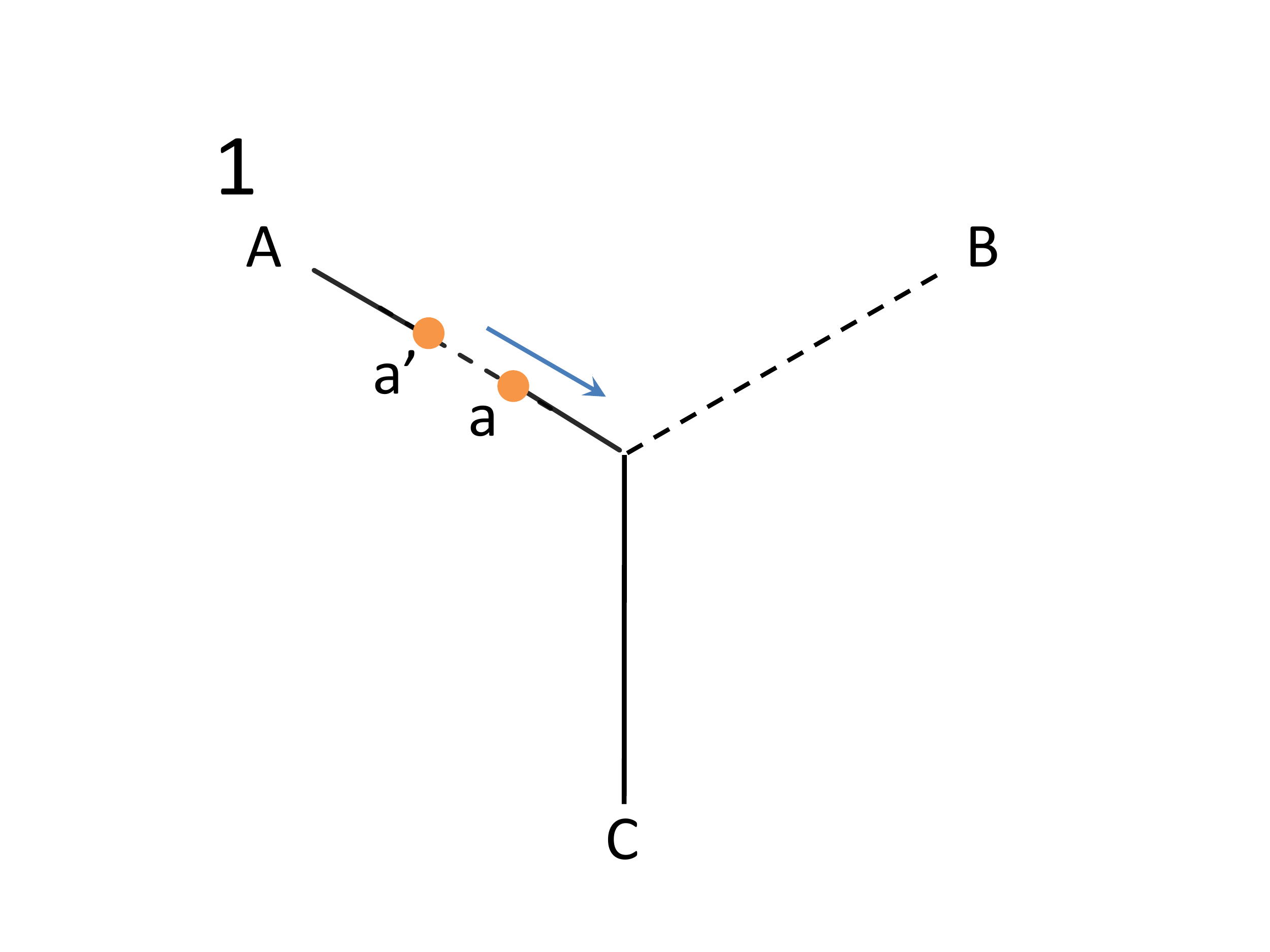}}
\subfigure{
\includegraphics[trim=4cm 1.5cm 4cm 2.5cm, clip, width=.45\columnwidth]{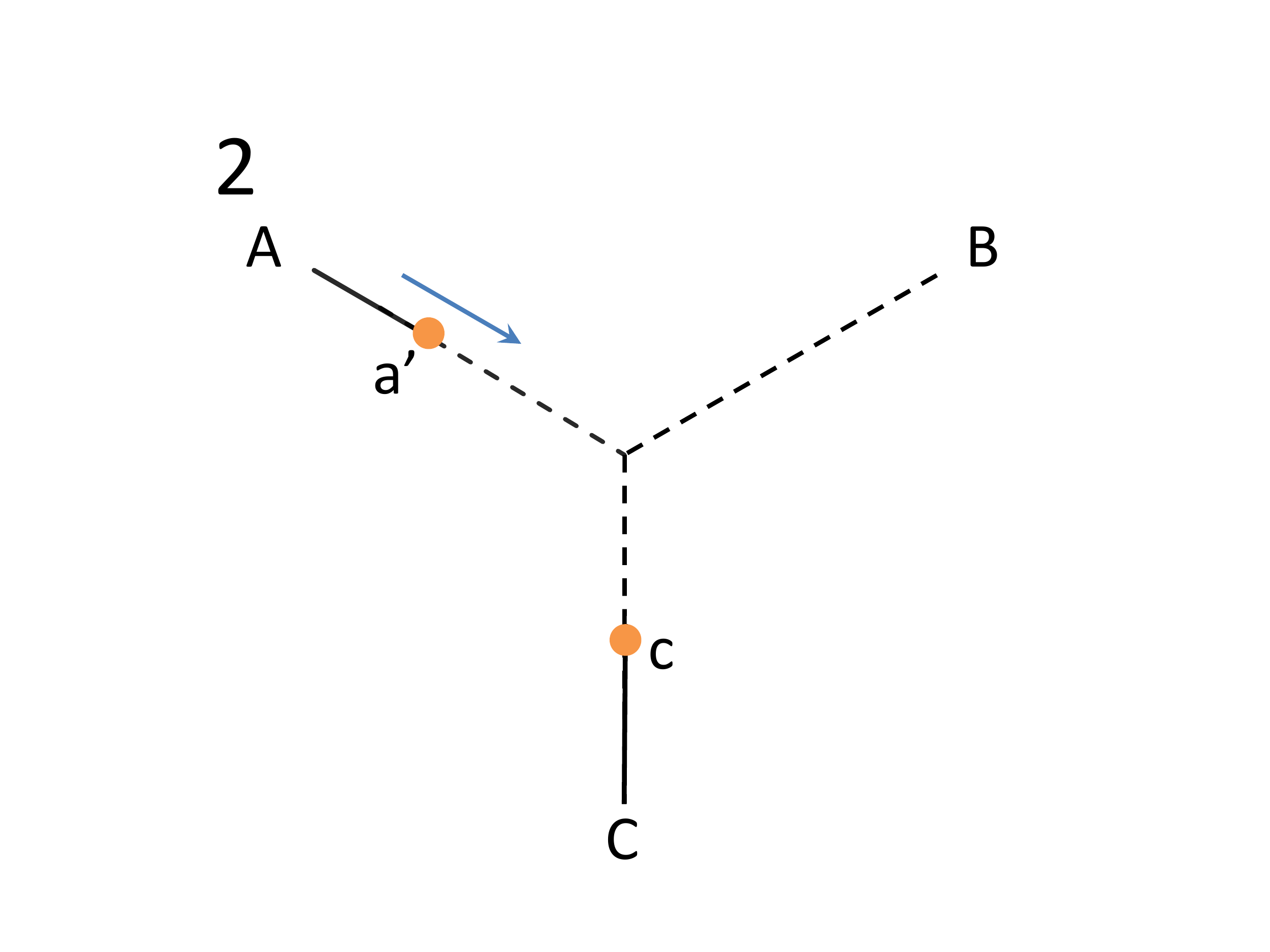}}
\subfigure{
\includegraphics[trim=4cm 1.5cm 4cm 2.5cm, clip, width=.45\columnwidth]{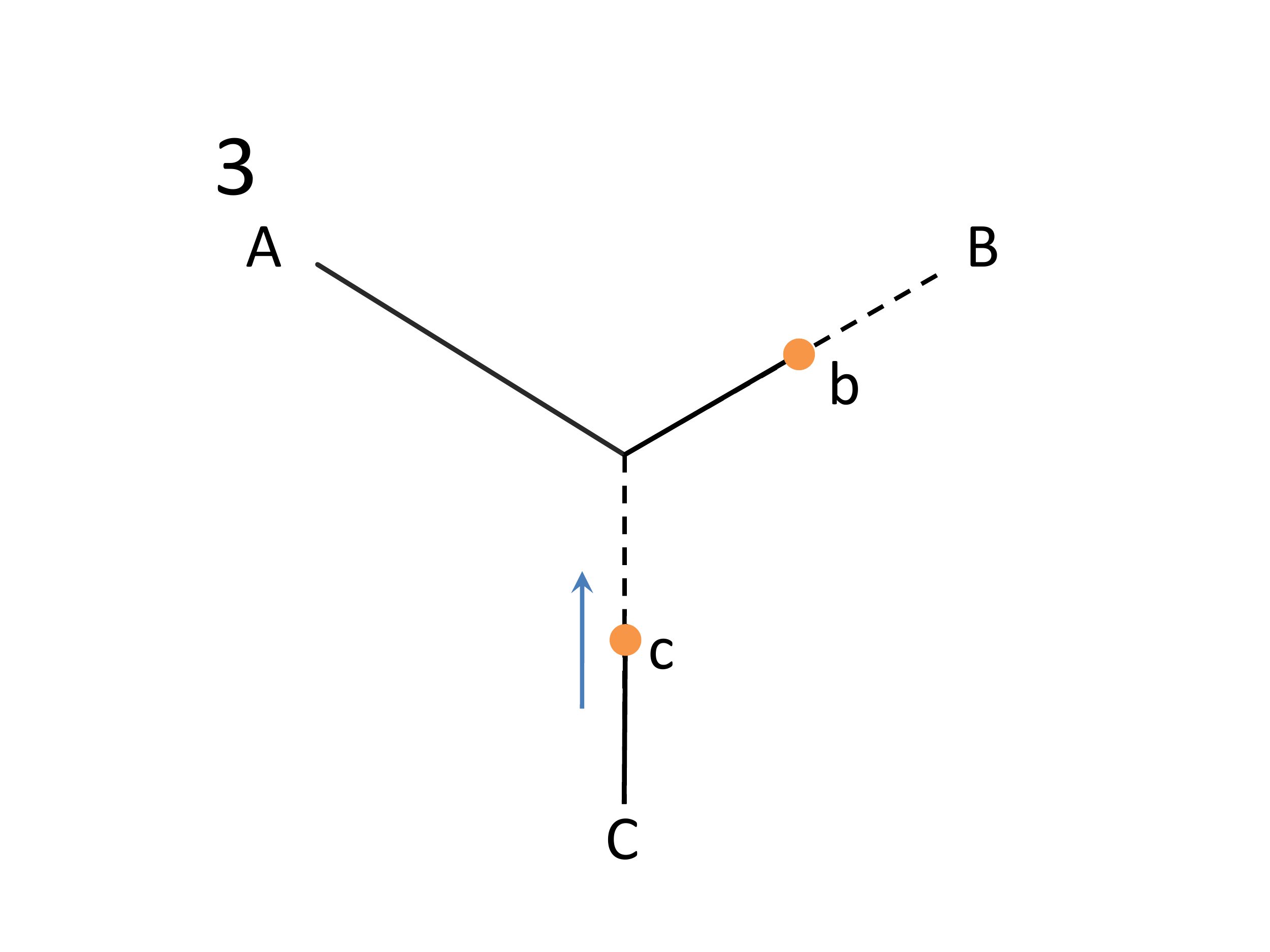}}
\subfigure{
\includegraphics[trim=4cm 1.5cm 4cm 2.5cm, clip, width=.45\columnwidth]{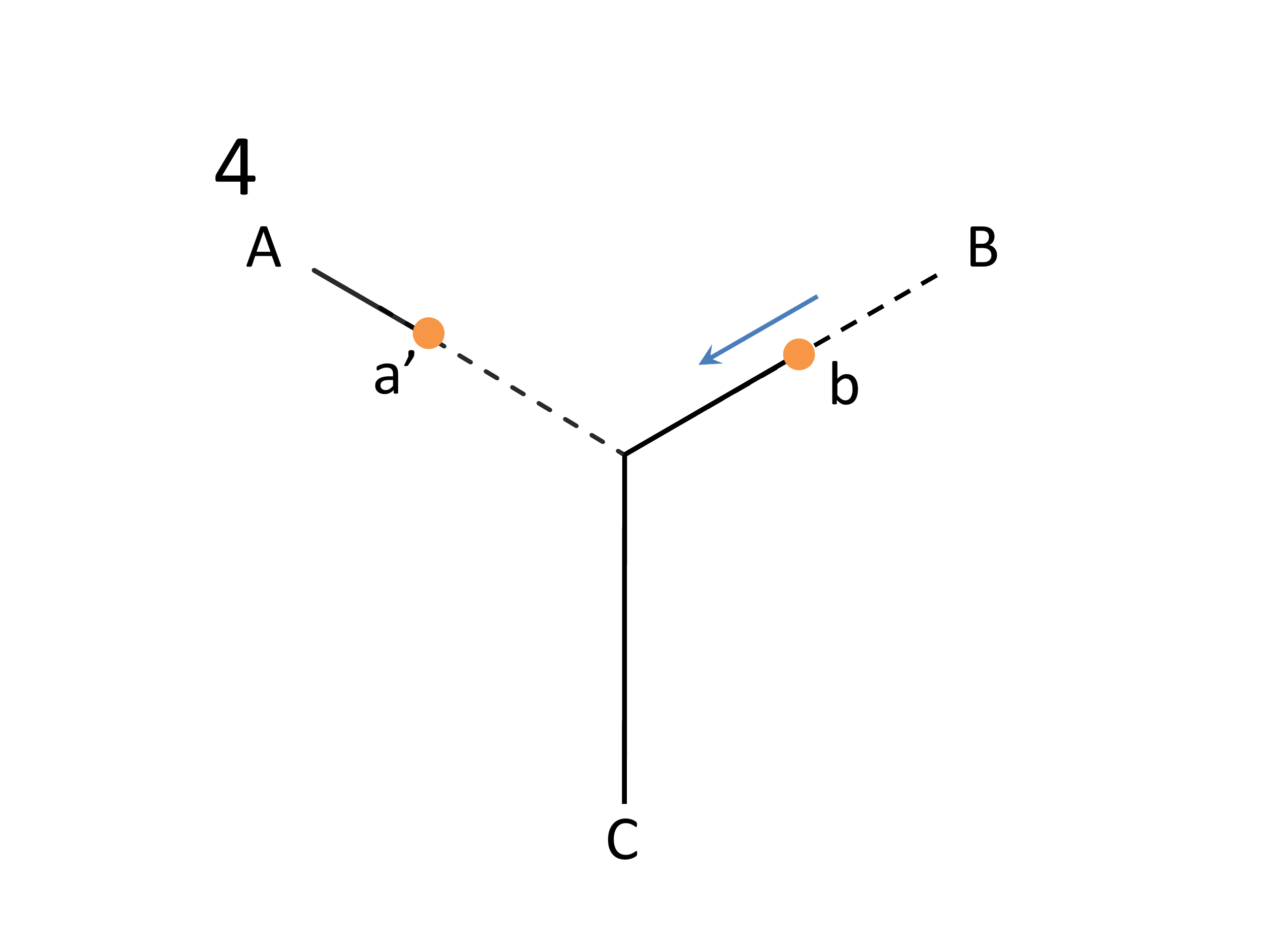}}
\subfigure{
\includegraphics[trim=4cm 1.5cm 4cm 2.5cm, clip, width=.45\columnwidth]{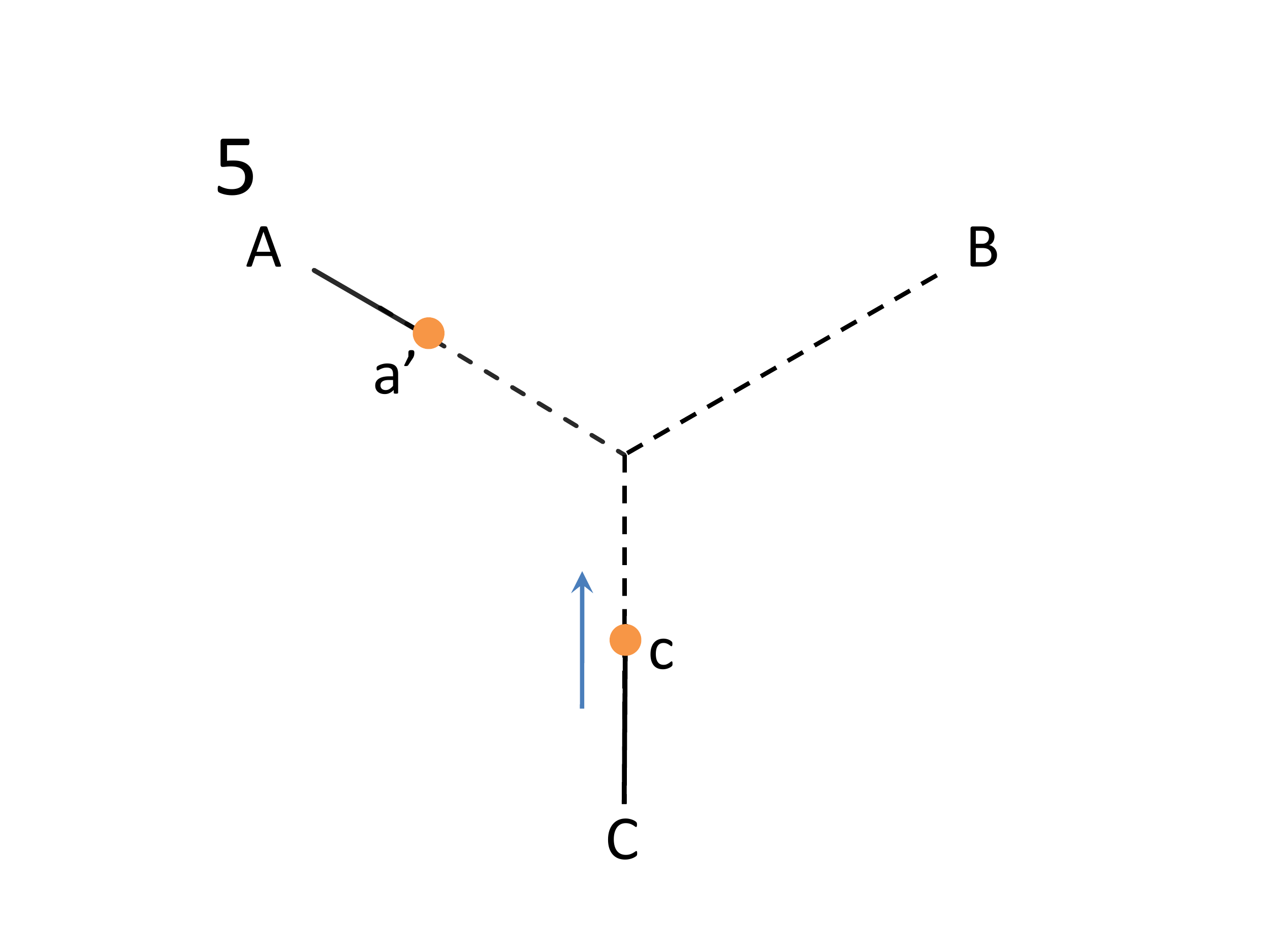}}
\subfigure{
\includegraphics[trim=4cm 1.5cm 4cm 2.5cm, clip, width=.45\columnwidth]{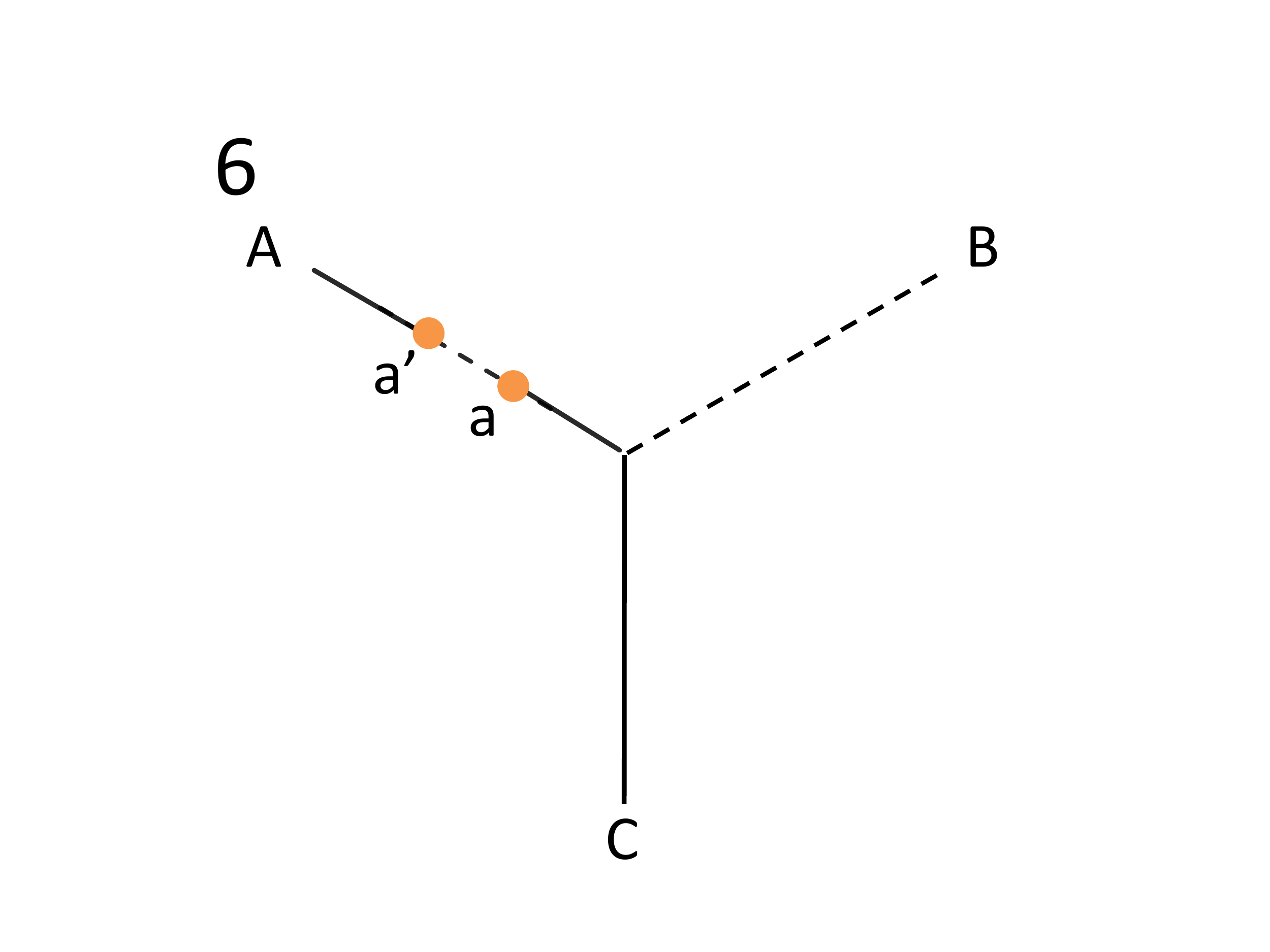}}
\caption{(Color Online) This series of figures shows the interchange of two Majorana bound states at the ends of a (dashed) non-topological region through the activation and deactivation of the wire segments surrounding the junction.}
\label{T-fig2}
\end{figure}
\subsection{Exchange of Majorana endpoints of separate Topological regions}
Suppose now that the two Majorana bound states that we wish to exchange begin as endpoints of different topological regions, as shown in Fig.~\ref{T-fig2}. We begin the exchange process by contracting the topological region from segment $A$ into segment $C$, so that one of the Majorana bound states moves from point $a$ to point $c$. We then move the remaining Majorana across the junction by expanding the other topological region from point $a'$ to point $b$ in segment $B$. This leads to
\begin{eqnarray}
\gamma_{c}(t_1)&=&\C_{ca}\gamma_{a}(0)\nonumber\\
\gamma_{b}(t_2)&=&\e_{ba'}\gamma_{a'}(0).
\end{eqnarray}
The time $t_i$ here is the time at which step $i$ in Fig.~\ref{T-fig2} is completed.
We are now faced with something of a dilemma: in order to move the Majorana from segment $C$ back into segment $A$, we must first extend the topological region from point $c$ to the junction, and then contract it into segment $A$. During this process, we bring three topological regions together at the junction, a step that was unnecessary in the previous type of exchange.

Due to the ambiguity in which of the three segments the junction point lies on, we cannot use our previously established $\C$ and $\e$ moves to carry out this process. Instead, we avoid the ambiguity by introducing a new factor $\mathcal{U}$ to describe this type of move (expand into junction and contract away). Then
\begin{equation}
\gamma_{a'}(t_3)=\mathcal{U}_{a'c}\gamma_{c}(t_1)
\end{equation}
The requirement of proper normalization for $\gamma_{a}$ and $\gamma_{c}$ still holds, so $\mathcal{U}_{a'c}=\pm1$. Reversibility of the process leads to $\mathcal{U}_{a'c}=\mathcal{U}_{ca'}$.

Continuing with the exchange, we contract the topological region from segment $B$ into segment $C$, then extend it back into segment $A$ to bring the remaining Majorana to point $a$. That is, \begin{equation}
\gamma_{a}(\tau=t_6)=\e_{ac}\C_{cb}\gamma_b(t_1)
\end{equation}
In total, we have
\begin{eqnarray}
\gamma_{a'}(\tau)&=&\mathcal{U}_{a'c}\C_{ca}\gamma_{a}(0)\nonumber\\
\gamma_{a}(\tau)&=&\e_{ac}\C_{cb}\e_{ba'}\gamma_{a'}(0).
\end{eqnarray}
By the argument of Sec.~\ref{sec-sign} and given Assumption 1, the two prefactors above must differ by a $-$ sign. That is, it must be that
\begin{equation}\label{eq-u}
\mathcal{U}_{a'c}=-\C_{cb}\e_{ba'}
\end{equation}
Because the points $a'$, $b$ and $c$ are arbitrary, this equation must hold whenever all three points are in different segments around the junction.

As we have above, we now separate the effect of the exchange process two factors, one describing motion along $A$ and one a property of the junction itself. Defining $\chi^n$ by
\begin{equation}\label{eq-chi-n}
\chi^n=v^n_a \e_{ab}\C_{bc}\e_{ca_-},
\end{equation}
we have that
\begin{eqnarray}\label{eq-exchange-n}
\gamma_{a'}(\tau)&=&\e_{a_-a'}v^n_a\chi^n\gamma_{a}\nonumber\\
\gamma_{a}(\tau)&=&-\e_{a_-a'}v^n_a\chi^n\gamma_{a'},
\end{eqnarray}
where we have used Eq.~(\ref{eq-vac-consistency2}) to permute indices where necessary. As in the case of Majoranas connected by a topological region, the value of $\chi$ is consistent when under transformations that rotate the three segments ($A\rightarrow B\rightarrow C$), and antisymmetric under the exchange of any two segments. As with $\chi^t$, $\chi^n$ is independent of the points chosen on the three segments for the representation given by Eq.~(\ref{eq-chi-n}).  Because of these properties, we shall refer to $\chi$ as the chirality of the junction, and show that $\chi=\chi^n=\chi^t$.

\begin{figure}
\subfigure{
\includegraphics[trim=4cm 1.5cm 4cm 2.5cm, clip, width=.45\columnwidth]{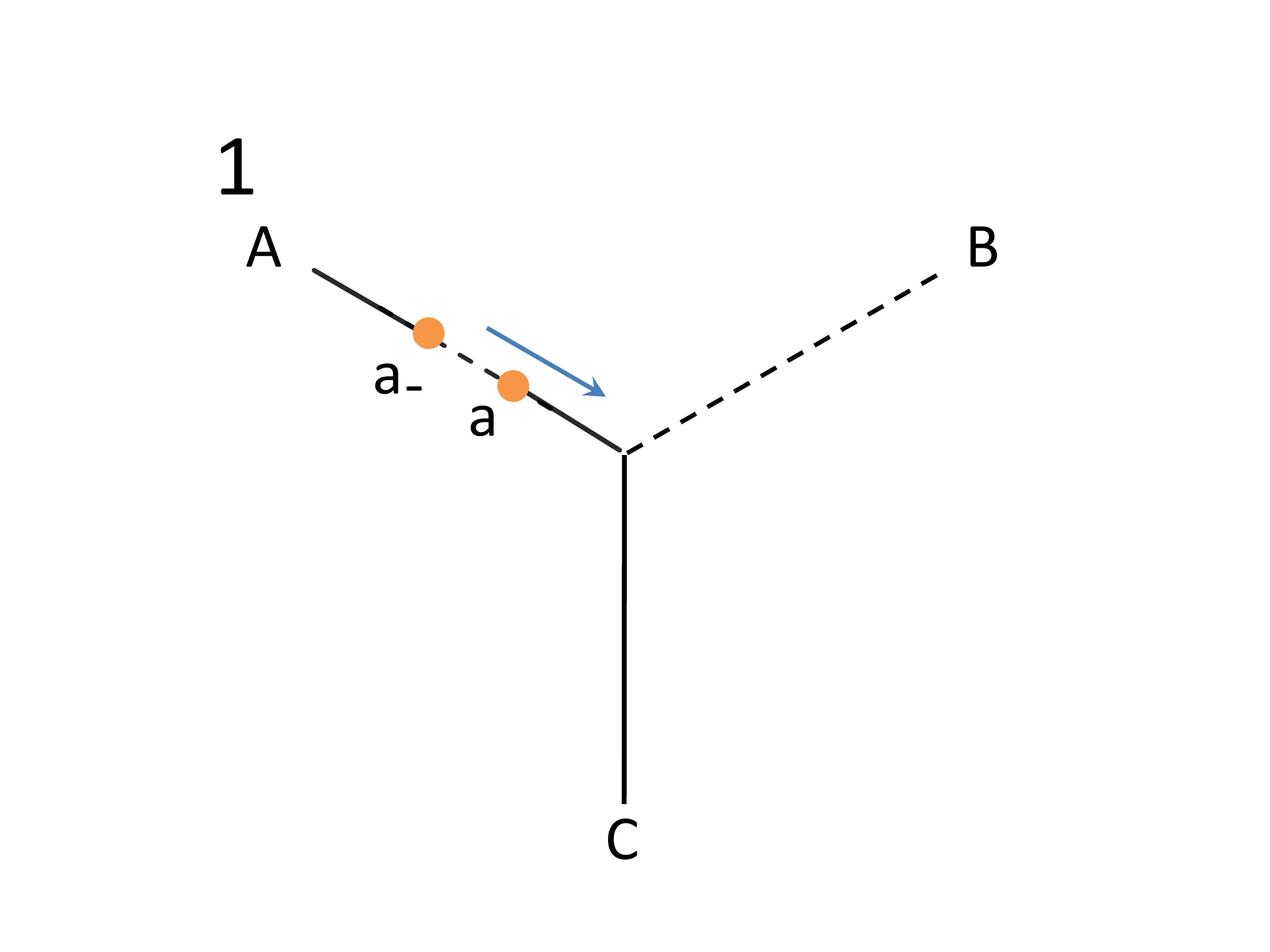}}
\subfigure{
\includegraphics[trim=4cm 1.5cm 4cm 2.5cm, clip, width=.45\columnwidth]{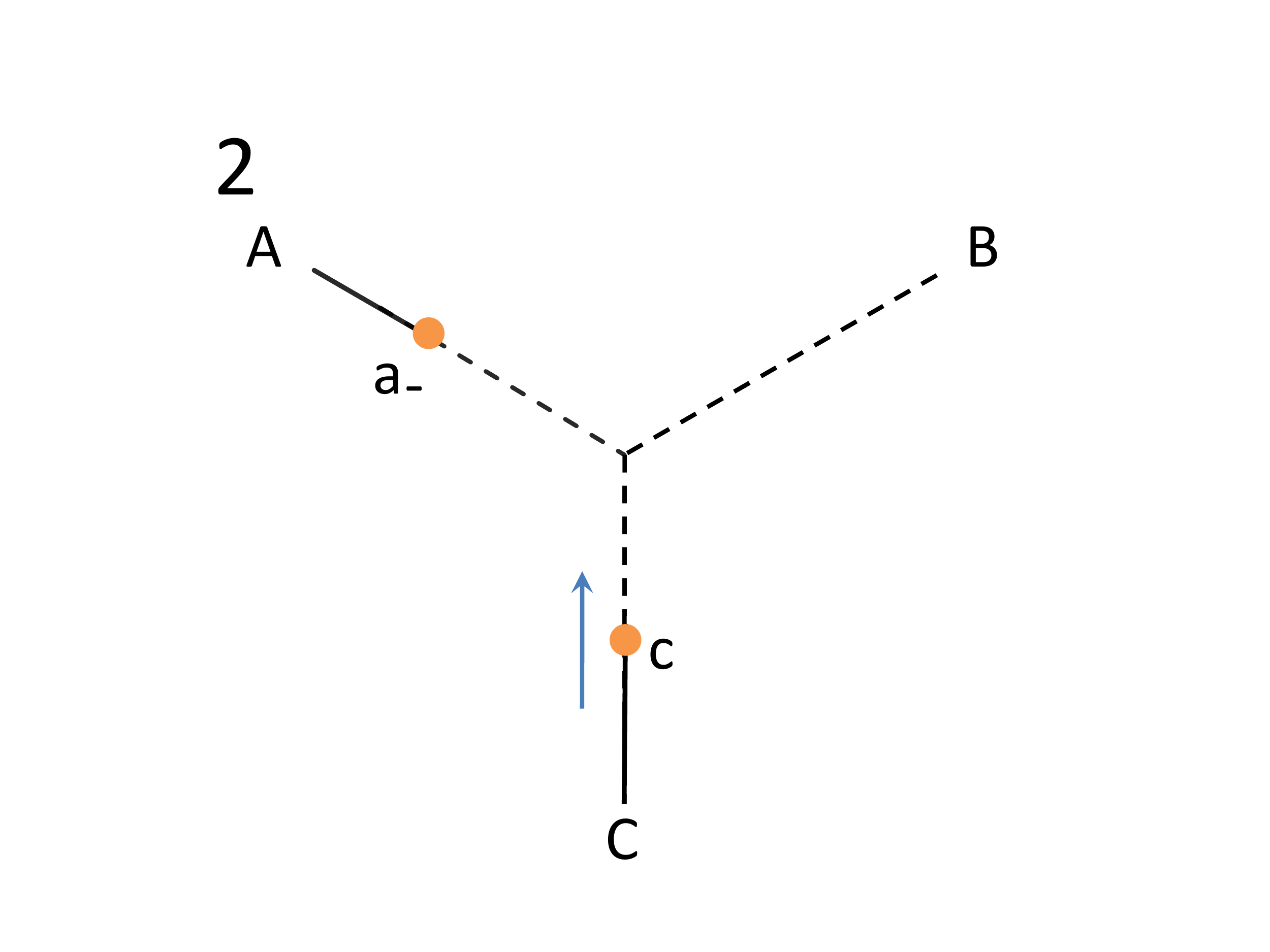}}
\subfigure{
\includegraphics[trim=4cm 1.5cm 4cm 2.5cm, clip, width=.45\columnwidth]{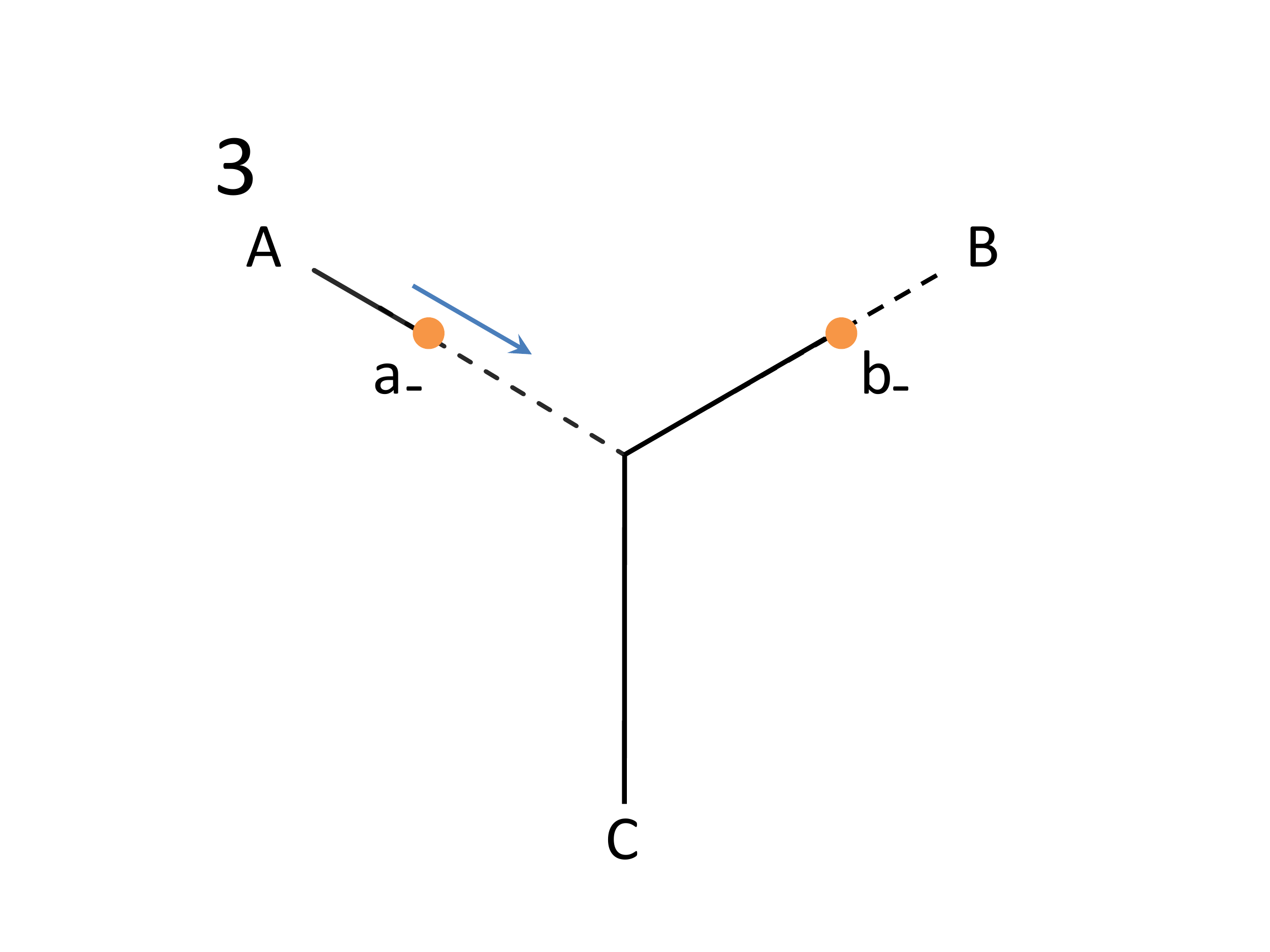}}
\subfigure{
\includegraphics[trim=4cm 1.5cm 4cm 2.5cm, clip, width=.45\columnwidth]{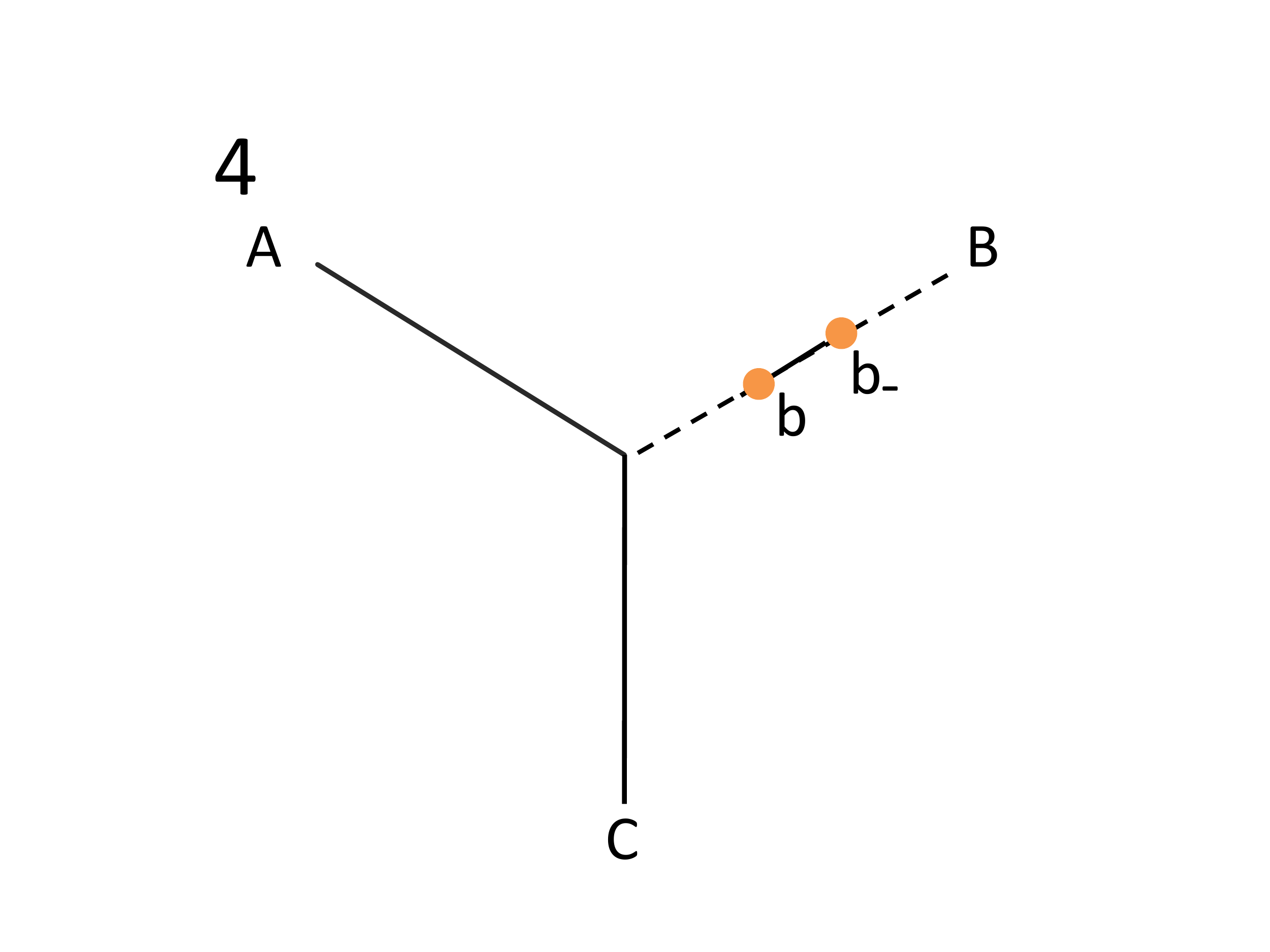}}
\caption{(Color Online) Majoranas at the endpoints of different topological regions are transformed into Majoranas at the endpoints of the same topological region by a three step process. The third step requires the use of the factor $\mathcal{U}_{ba_-}$ as in Eq.~(\ref{eq-u}).}
\label{fig-umove}
\end{figure}

In order to see this, it is instructive to consider another process, in which two Majoranas connected by a non-topological region are transformed into two Majoranas connected by a topological region (Fig.~\ref{fig-umove}). It is shown by this process, as well as by the arguments leading to Eqs.~(\ref{eq-vac-consistency}) and (\ref{eq-vac-consistency2}) that
\begin{equation}
\mathcal{U}_{ba_-}=-v_a^nv_b^t e_{b_-c}c_{ca}.
\end{equation}
When combined with Eq.~(\ref{eq-u}), this gives
\begin{equation}
\C_{bc}\e_{ca_-}=v_a^nv_b^t\e_{b_-c}\C_{ca}.
\end{equation}
Using this with Eq.~(\ref{eq-chi-consistent}) and Eq.~(\ref{eq-chi-n}) shows that $\chi^n=\chi^t$. That is, the effective junction chirality is independent of whether the Majoranas involved in the exchange are endpoints of the same or different topological regions.
\subsection{Braid transformation}\label{sec-no-ambiguity}
We can combine the two cases for Majorana braiding into a single notation, writing the unitary time evolution operator for the entire process as
\begin{equation}
U_\chi(\tau)=e^{\frac{\pi}{4}\mathcal{M}_{a_-a'}v_a\chi\gamma_{a'}\gamma_{a}},
\end{equation}
where $\mathcal{M}$ is the sign (either $\C$ or $\e$) required to move the Majorana at $a'$ to the point $a_-$ close to $a$. $v_a$ is the vacuum channel as determined by the local Hamiltonian of the form (\ref{ham}).

This operator has the advantage of invariance under the sign ambiguity inherent in the Majorana fermion description. While the time-evolution of a Majorana fermion
leads to a definite result $\gamma(\tau)=U_\chi^\dagger(\tau) \gamma(0) U_\chi(\tau)$
for, it is important to remember the following
caveat about the definition of the Majorana operators themselves.
 If $\gamma(t)$ is a
 zero energy operator (in the sense that $[H_{BCS}(t),\gamma_j(t)]=0$), then so is $\tilde{\gamma}_j(t)=\nu\gamma(t)$ for $\nu=\pm 1$.
This is essentially a phase ambiguity in the definition of a Majorana fermion (note that
a general phase $e^{i \theta}$ is not allowed since $\gamma^\dagger(t)=\gamma(t)$). However, in redefining any particular Majorana operator involved in the exchange, one must also change the sign of any vacuum channel, contraction factor or expansion factor in which that Majorana is involved. For example, in redefining $\gamma_a\rightarrow-\gamma_a$, one must also change $v_a^{(t,n)}\rightarrow-v_a^{(t,n)}$, $\C_{xa}\rightarrow-\C_{xa}$ and $\e_{xa}\rightarrow-\e_{xa}$ in order to maintain consistency with the definition of other Majorana operators. Because the definition of $\chi^n$ (or $\chi^t$) contains factors involving each Majorana operator twice, $\chi$ is invariant under this transformation. Likewise, if one changed $\gamma_a'\rightarrow-\gamma_a'$, $U_\chi$ as defined above remains invariant because $\mathcal{M}_{a_-a'}$ would also change sign.
\section{Discussion}
\subsection{Comparison of Junction Chiralities}\label{sec-2-junction}
It is important to note that the chirality of a given junction is not set a priori by the general analysis considered here. Instead, it is set by the underlying microscopic parameters and the exact method by which Majoranas are moved through the junction. In particular, a junction's chirality may be altered by a local defect that binds a fermion in only one of the two phases. Such a defect causes the Majorana bound states to emit or absorb a fermion whenever they pass by the defect, changing the sign of each Majorana. If the defect is located at the junction, it will be passed three times by the Majoranas during an exchange. Each Majorana would therefore acquire an additional minus sign during the exchange, changing the effective chirality of the junction.\endnote{ We detail a particular method of moving the Majoranas that does not suffer from this problem in a companion paper (Ref.~\onlinecite{Paper 2}).}

In any case, different junctions within a wire network do not necessarily have the same chirality. To understand the effect that this might have on a computational algorithm, consider the situation in which a pair of Majorana fermions $\gamma_2$ and $\gamma_3$ are exchanged clockwise via first one junction with $\chi=\chi_1$ then through a second junction with $\chi=\chi_2$. If $\chi_1\neq\chi_2$, one of these exchanges is effectively a counter-clockwise, rather than a clockwise, braid. In this case, the net effect is no braid at all, rather than a double braid.

More explicitly, we shall examine the effect of this manipulation when $\gamma_2$ forms a qubit with $\gamma_1$, and $\gamma_3$ forms a qubit with $\gamma_4$. Then the fusion channel of $\gamma_1$ and $\gamma_2$ evolves as
\begin{equation}
U^\dagger_{\chi_1} U^\dagger_{\chi_2} i\gamma_1\gamma_2 U_{\chi_1}U_{\chi_2}=-\chi_1\chi_2i\gamma_1\gamma_2
\end{equation}
That is, the $\gamma_1$, $\gamma_2$ qubit flips do to this double exchange only if the chiralities of the two junctions being used are the same. Likewise, the $\gamma_3$, $\gamma_4$ qubit also flips only if $\chi_1=\chi_2$. Along with a method of measuring the qubit states (such as the fractional Josephson effect as propose in Ref.~\onlinecite{alicea1}), this double-exchange test may be used to determine the relative chirality of different junctions in a wire network.
\subsection{Multiply connected networks}\label{sec-multiply}
It is worth noting that while the result of a braiding of two Majoranas is uniquely determined by the properties of the underlying wire network, the braid result may be dependent on the path taken through the wire system. Consider, for instance, the possibility that segments B and C in the T-junction are connected via a loop of wire. It would then be possible to interchange the Majoranas at positions $a$ and $a'$ in Fig.~\ref{T-fig} clockwise by taking them from segment A to segment B then transporting them around the loop to segment C, and finally back through the junction into A. The result of this transportation is
\begin{eqnarray}
\gamma_{a'}(\tau)&=&\e_{a'c}\widetilde{\e}_{cb'}\e_{b'a}\gamma_{a}(0)=\C_{a_-a'}v^t_a\chi\mathcal{L}\gamma_{a}(0)\nonumber\\
\gamma_{a}(\tau)&=&\C_{ac'}\widetilde{\C}_{c'b}\C_{ba'}\gamma_{a'}(0)=-\C_{a_-a'}v^t_a\chi\mathcal{L}\gamma_{a}(0),
\end{eqnarray}
where  $\mathcal{L}=-\widetilde{\e}_{cb}\e_{bc}=-\widetilde{\C}_{cb}\C_{bc}$, and where the $\widetilde{\e}_{cb'}$ indicates expansion around the loop from $b$ to $c$, rather than through the junction. That $\widetilde{\e}_{cb}\e_{bc}=\widetilde{\C}_{cb}\C_{bc}$ may be shown using Assumption 2, Eq.~(\ref{eq-expansion-consistency}) and Eq.~(\ref{eq-vac-consistency}). The result here differs from Eq.~(\ref{eq-exchange-t}) by the 'loop factor' $\mathcal{L}$. This factor is independent of the points $b$ and $c$ used in the definition, as well as being even under the exchange of segments $B$ and $C$. $\mathcal{L}=-1$ may be thought of as indicating the existence of a flux (real or effective) through the loop in the network.

\begin{figure}
\subfigure{
\includegraphics[trim=4cm 2cm 4cm 3cm, clip, width=.4\columnwidth]{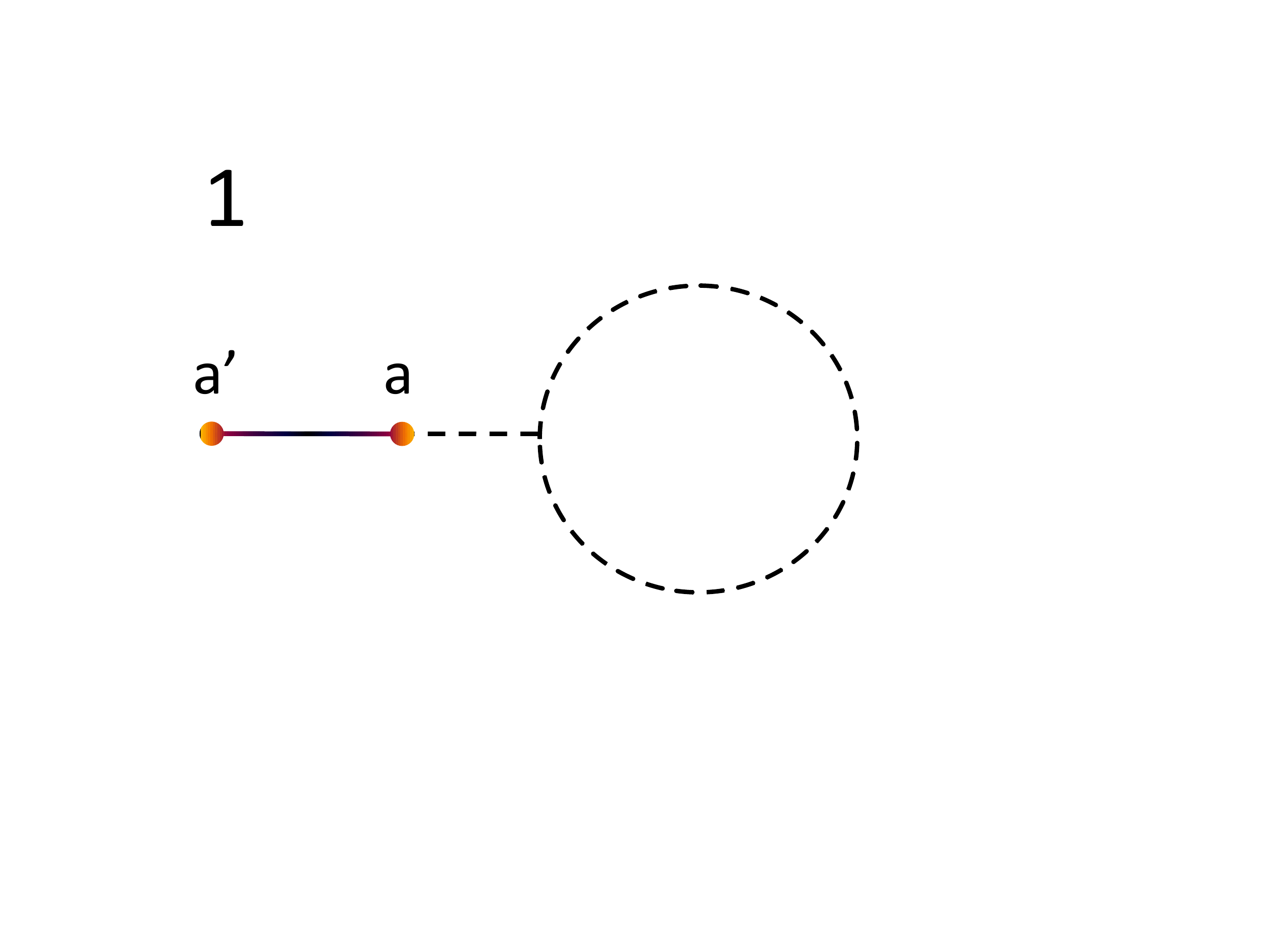}}
\subfigure{
\includegraphics[trim=4cm 2cm 4cm 3cm, clip, width=.4\columnwidth]{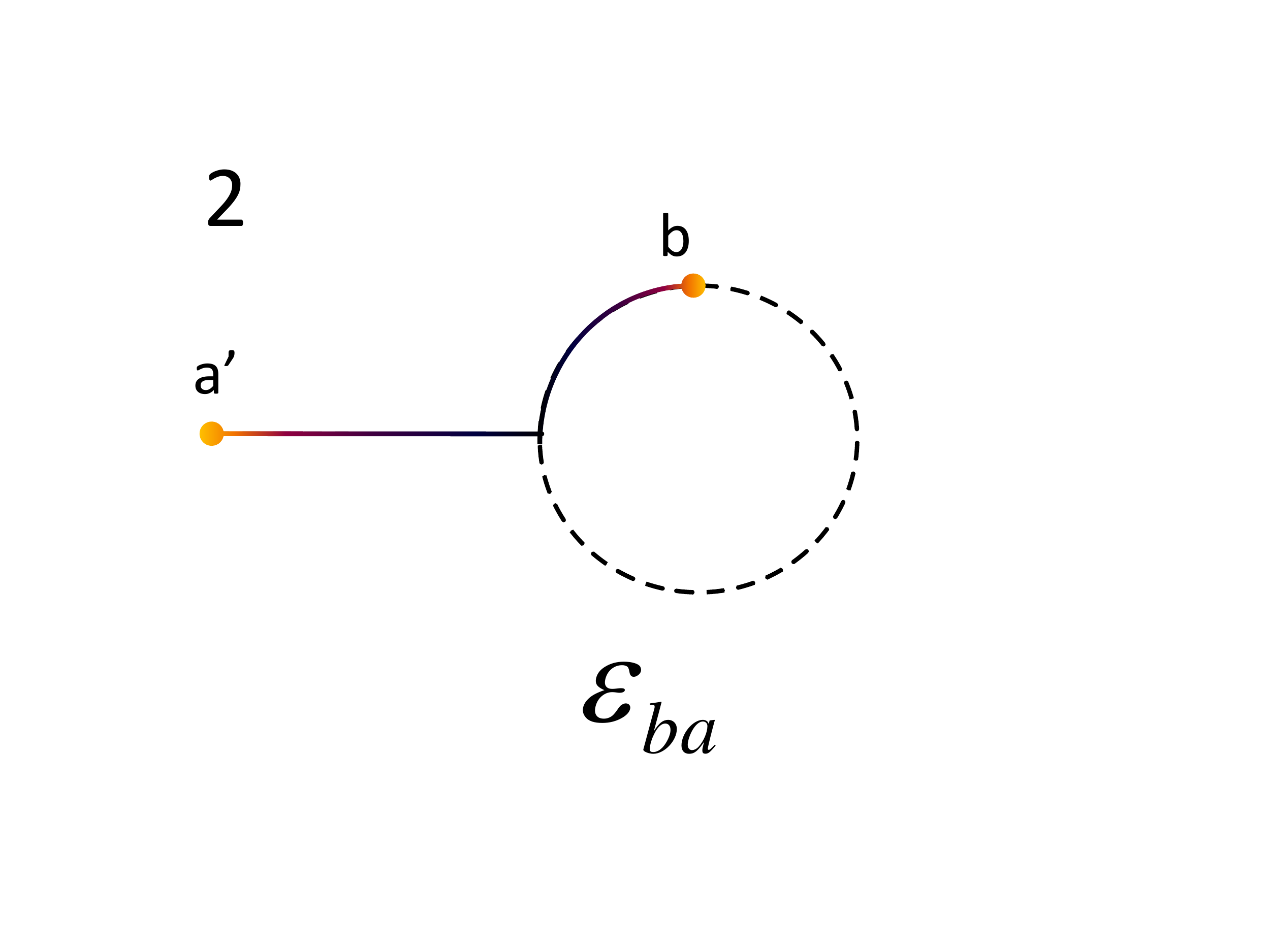}}
\subfigure{
\includegraphics[trim=4cm 2cm 4cm 3cm, clip, width=.4\columnwidth]{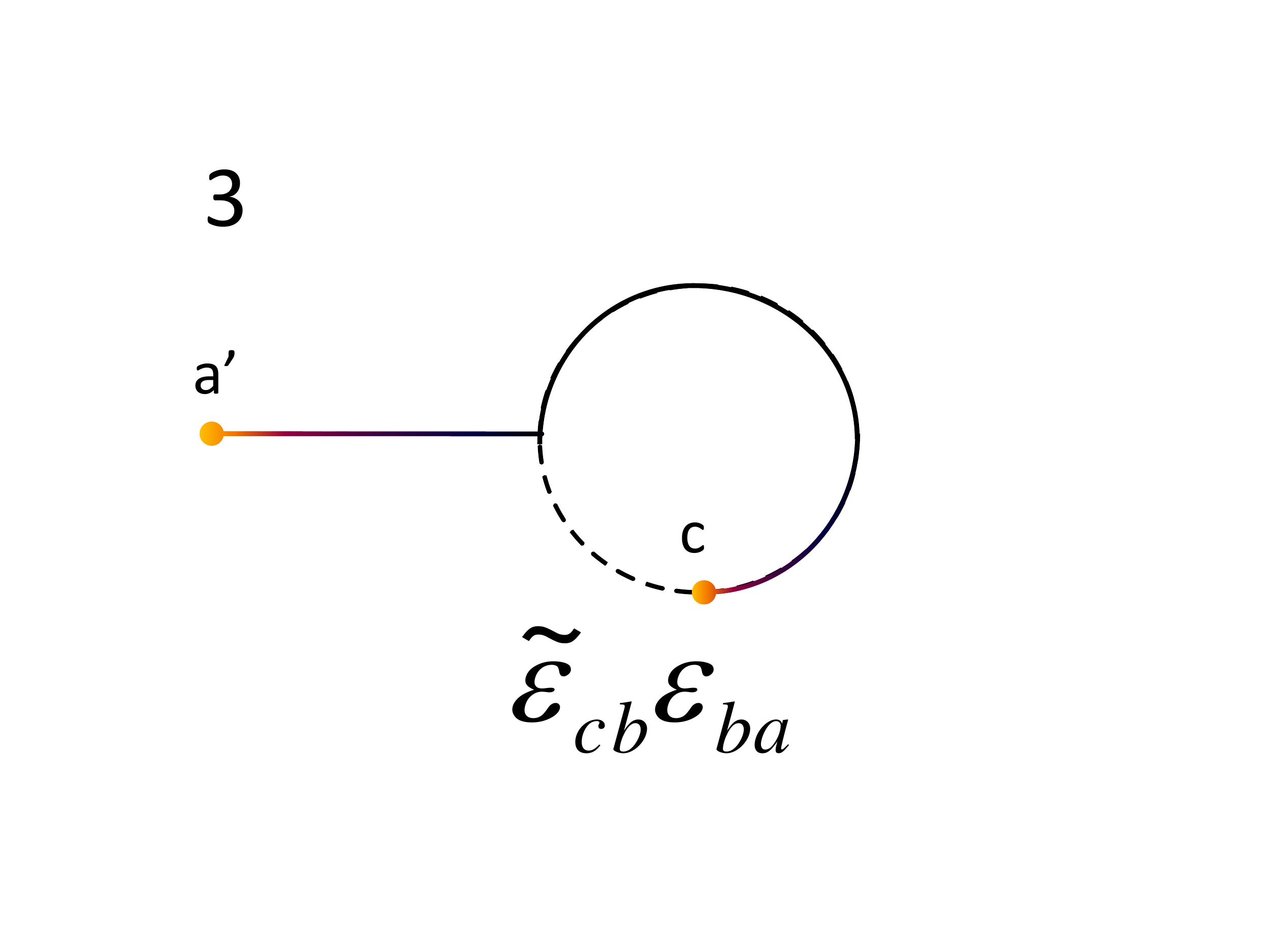}}
\subfigure{
\includegraphics[trim=4cm 2cm 4cm 3cm, clip, width=.4\columnwidth]{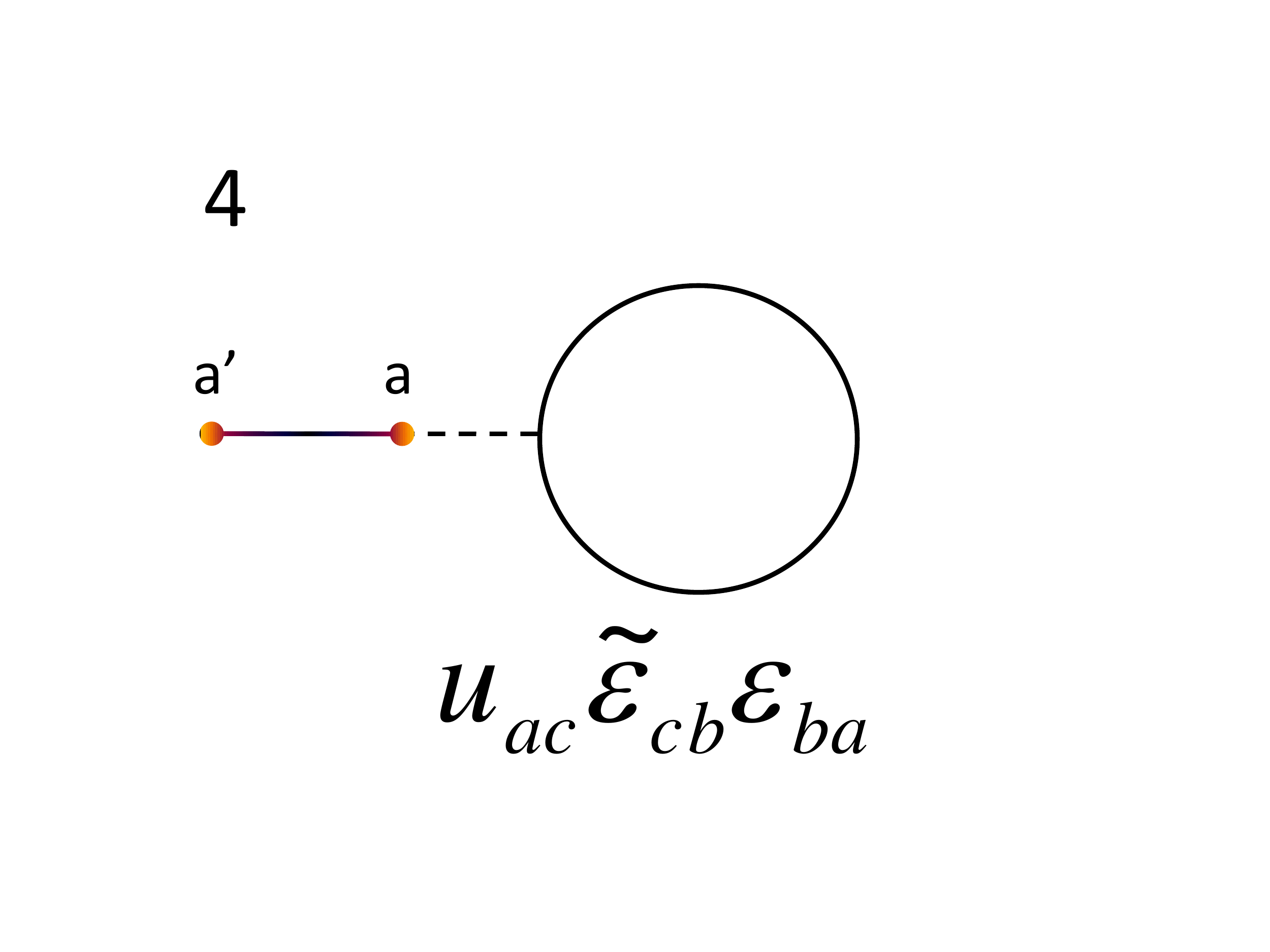}}

\caption{(Color Online) This series of figures shows the transport of a Majorana fermion around a loop in a multiply connected wire network. Each figure indicates the total phase factor acquired in the motion of the Majorana to that point. Since $\mathcal{U}_{ac}=-\C_{cb}\e_{ba}$ the total phase factor acquired by the Majorana is $\mathcal{L}=-\widetilde{\e}_{cb}\e_{bc}$.}
\label{fig-loop}
\end{figure}

Consider Majorana fermions at points $a'$ and $a$ near a loop in the wire network as pictured in Fig.~\ref{fig-loop}. Transporting the Majorana around the loop and back to its original position yields a transformation
\begin{equation}
\gamma_{a}(\tau)=\mathcal{L}\gamma_{a}(0).
\end{equation}
Since $\gamma_{a'}$ is unaffected by this process, if $\mathcal{L}=-1$  the fermion parity operator in the $(a',a)$ (proportional to $\imath\gamma_{a'}\gamma_{a}$) changes sign. This indicates that if there were an odd number of fermions in the $(a',a)$ region before the process began, then there will be an even number afterward and vice versa. This may appear to contradict the conclusions of Sec.~\ref{sec-proof}; however, the conclusion that the fermion parity is conserved in the motion of Majorana fermions was dependent upon the assumption that $H_{BCS}(\tau)=H_{BCS}(0)$, which is not the case here. In fact, the loop has switched phase. The extra fermion parity may therefore be found in the loop, either in a localized bound state or in the alteration of the ground state of the loop from even to odd parity if a real external flux is present. In this sense, $\mathcal{L}$ is seen to indicate the presence or absence of an effective flux through the loop.

\subsection{Applicability}
The results of this paper apply to any one-dimensional wire system supporting Majorana fermions at the boundaries between two phases for the system, so long as the assumptions laid out in Sec.~\ref{sec-clockwise-exchange} hold true. The details of the underlying implementation of such a system, e.g. in semiconductor/superconductor heterostructures with large spin-orbit coupling\cite{alicea1}, will determine the values of the factors $\chi$ and $\mathcal{L}$ that we have identified here for junctions and loops, respectively. Our results establish a concrete framework for describing and tracking the non-Abelian transformations made by moving Majorana fermions through a quantum wire network.

\acknowledgements{We thank Parsa Bonderson and Anton Akhmerov for helpful discussions. We are grateful to the Aspen Center for Physics for hospitality during the 2010 summer program \emph{Low Dimensional Topological Systems}.  D.J.C. is supported in part by the DARPA-QuEST program. S.T. acknowledges support from DARPA-MTO Grant No: FA 9550-10-1-0497. J.D.S. is supported by DARPA-QuEST, JQI-NSF-PFC, and LPS-NSA.}

\end{document}